\documentclass[prd,nofootinbib]{revtex4}
\usepackage{amsmath}
\usepackage{graphicx}
\usepackage{amssymb}

\begin{document}

\title{Spinning compact binary inspiral II: Conservative angular dynamics}
\author{L\'{a}szl\'{o} \'{A}rp\'{a}d Gergely$^{1,2\star }$}
\affiliation{$^{1}$Department of Theoretical Physics, University of Szeged, Tisza Lajos
krt 84-86, Szeged 6720, Hungary\\
$^{2}$Department of Experimental Physics, University of Szeged, D\'{o}m t%
\'{e}r 9, Szeged 6720, Hungary\\
{\small {$^\star$ E-mail: gergely@physx.u-szeged.hu\qquad } }}

\begin{abstract}
We establish the evolution equations of the set of independent variables
characterizing the 2PN rigorous conservative dynamics of a spinning compact
binary, with the inclusion of the leading order spin-orbit, spin-spin and
mass quadrupole - mass monopole effects, for generic (noncircular,
nonspherical) orbits. More specifically, we give a \textit{closed system of
first order ordinary differential equations} for the orbital elements of the
osculating ellipse and for the angles characterizing the spin orientations
with respect to the osculating orbit.

We also prove that (i) the relative angle of the spins stays constant for
equal mass black holes, irrespective of their orientation, and (ii) the
special configuration of equal mass black holes with equal, but antialigned
spins, both laying in the plane of motion (leading to the largest recoil
found in numerical simulations) is preserved at 2PN level of accuracy, with
leading order spin-orbit, spin-spin and mass quadrupolar contributions
included.
\end{abstract}

\date{\today }
\maketitle

\section{Introduction}

Compact binaries composed of neutron stars or stellar size black holes are
among the most likely sources to emit gravitational waves in the frequency
range of the Earth-based gravitational wave detectors LIGO and Virgo \cite%
{LIGO}. Supermassive black holes in the mass range of $3\times 10^{6}\div
3\times 10^{9}$ solar masses reside in the centers of galaxies and following
the merger of their host galaxies, they also merge. In the process they
create powerful gravitational waves, detectable in the lower mass range by
the space mission LISA \cite{LISA}.

By definition the \textit{inspiral} is the regime of the orbital evolution,
during which the post-Newtonian (PN) parameter $\varepsilon
=Gm/c^{2}r\approx \left( v/c\right) ^{2}$ (where $m\equiv m_{1}+m_{2}$ is
the total mass, $r$ and $v$\thinspace\ the orbital separation and relative
velocity of the binary) is small and where the leading order dissipation is
due to the gravitational waves. As two galaxies merge, their supermassive
black holes are subject to both gravitational radiation and dynamical
friction. The former overcomes the latter at about $\varepsilon
_{in}=10^{-3} $ (the actual number only weakly depends on both the stellar
distribution and mass \cite{SpinFlip}). During the inspiral which follows,
the parameter $\varepsilon $ increases. When $\varepsilon $ approaches its
value at the innermost stable orbit, the PN description becomes increasingly
nonaccurate, therefore the subsequent \textit{plunge} is better described by
numerical evolutions, or as an alternative, by expressions traced back to
the PN approach, arising either from the effective one-body model,
calibrated to numerical relativity simulations \cite{EOB} or from a
phenomenological transition phase, with coefficients again calibrated by
comparison with specific, numerically generated waveforms \cite{plunge}.
Finally, the \textit{ringdown} follows, when the newly formed black hole
radiates away its physical characteristics, with the exception of mass, spin
and possibly electric charge (for a review of quasinormal modes of black
holes see Ref. \cite{ringdown}).

The spin and quadrupole moment of the supermassive black hole at the
Galactic center can be measured via astrometric monitoring of stars orbiting
at milliparsec distances \cite{no-hair-SMBH-test}, and this can also be a
test of the general relativistic no-hair theorem.

The spin affects the horizon of the black holes, therefore those
observations on black holes which indicate the size of the horizon, will
also lead to indirect spin \textit{magnitude} estimates. (Estimating the
quadrupole moment from observing a two-dimensional projection of the horizon
would be less straigtforward.) Both stellar size and supermassive black
holes can have accretion disks and jets in their active periods. Whenever
these observations are connected to the presence of a jet and the direction
of it can be identified (projection effects may again obstruct this), we
also obtain information on the black hole spin \textit{axis}, assuming it is
aligned with with the symmetry axis of the magnetic field and hence the jet
direction. Jets from rotating black holes have been shown to be stable \cite%
{McKinneyBlandford}. Spin direction can be also inferred from observations
on the radiation of the accretion disk.

Such observational spin estimates can be made at least by four methods:

i) Reverberation mapping of the observed optical/X-ray lines (highly excited
Mg, O, C) in active galaxies to determine the radius and velocity pattern of
the Broad Line Region. This depends on the metric, decreasing with
increasing spin. From such considerations the mass, spin and spin
orientation of the black holes can be estimated \cite{reverberation}. In
particular, information on the spin direction of the central black hole of
the Seyfert galaxy Mrk 110 was obtained by estimating the central black hole
mass in two independent ways. First, assuming that the broad emission lines
are generated in gas clouds orbiting within an accretion disk, the mass
could be determined as function of the inclination angle of the accretion
disk. Secondly, detecting the gravitational redshifted emission in the
variable fraction of the broad emission lines, a central black hole mass,
which is independent on the orientation of the accretion disk could be
deduced \cite{reverberation2}.

ii) There is a strong effort towards imaging with millimeter Very Long
Baseline Interferometry (VLBI) the event horizons of Sagittarius A* (SgrA*)
and Virgo A (M87), which again depend on the spin. For SgrA*, the radio
source at the center of our Milky Way, millimeter and infrared observations
require the existence of a horizon \cite{SagAhorVLBI}. Analysing the peaks
of the power density spectra in the light curves of X-ray flares from the
Galactic Center black hole the mass and spin were inferred \cite%
{SagAhorVLBI2}. A compact emission region (bright spot) in a circular orbit
and the lightcurves of its observed flux and polarisation depend on the mass
and spin of the black hole. The emitted polarisation fraction is polarised
orthogonally to the spin axis of the black hole \cite{SagAhorVLBI3}. Unlike
SgrA*, M87 exhibits a powerful radio jet, allowing future VLBI data to
constrain the size of the jet footprint, the jet collimation rate, and the
black hole spin \cite{M87horVLBI}.

iii) The size of the jet launching region in Active Galactic Nuclei (AGN) is
determined by the Blandford-Znajek effect, which in turn depends on the spin 
\cite{jetspin}. Measurement of the diameter of the jet base (e.g. in M87)
gives evidence for small sizes, regarded as signature of a large spin \cite%
{jetsize}.

iv) The low-energy cutoff in the energetic electron spectra of the jets
suggested by the radio spectra \cite{cutoff} is conveniently explained by
the pion decay resulting from proton-proton collisions \cite{piondecay}. The
latter mechanism require a relativistic temperature in the accretion disk
near the foot of the jet, which translates to the central black hole
spinning extremely fast \cite{HtempFspin}.

Some of these methods will certainly also work for stellar size black holes
due to scale invariance arguments in accretion phenomena. The jet/disk
geometry has been constrained for the stellar black holes XTE J1118+480 and
GX 339-4 \cite{Markoff}.

From all these observations we conclude that it is necessary to include the
spin and quadrupole moment of black holes when modeling their binary systems.

In this paper we investigate the 2PN rigorous conservative dynamics during
the inspiral of a spinning compact binary system, by including leading order
spin-orbit (SO), spin-spin (SS) and mass quadrupole - mass monopole (QM)
effects, for generic (noncircular, nonspherical) orbits. Because of these
interactions the spins undergo a precessional motion \cite{BOC}-\cite{BOC2}.
Various aspects related to the leading order contribution to both the
conservative and dissipative part of the dynamics due to the SO interaction
were discussed in Refs. \cite{KWW}-\cite{Kidder}, while the corrections
represented by the SS coupling in Refs. \cite{Kidder}-\cite{SS}, and by the
QM coupling in Refs. \cite{Poisson}-\cite{QM}. The radial motion under the
Newton-Wigner-Pryce spin supplementary condition \cite{NWP}, with all these
contributions included is fully solved in Ref. \cite{Kepler}. The
Hamiltonian approach including spins has been also widely discussed \cite%
{Ham}. Based on numerical work, empirical formulas for the final spin have
been proposed in Refs. \cite{finalspin}. Zoom-whirl orbits, which were known
to exist for particles orbiting Kerr black holes \cite{zwKerr}, also appear
in the framework of the PN formalism \cite{zwPN}, their likeliness
increasing with the spin \cite{zwNumeric}. Gravitational wave emission can
lead to a spin-flip \cite{LeahyWilliams}-\cite{Xspinflip} in X-shaped radio
galaxies \cite{LeahyWilliams}, \cite{Xshape}. It is a more recent result
that for mass ratios $0.3\div 0.03$ the combined effect of SO precession and
gravitational radiation will result in a substantial spin-flip already in
the inspiral phase \cite{SpinFlip}, \cite{SpinFlip2}.

With the spins and mass quadrupole moments included, the number of variables
in the configuration space increases drastically. In Ref. \cite{IndepVar} a 
\textit{minimal and conveniently chosen set of independent variables} for
such a system was established. Notational correspondence of some of these
variables with quantities employed in Refs. \cite{FC}-\cite{ABFO}, where the
dynamics of spinning compact binaries (without mass quadrupolar
contributions) has been also discussed, is established in Appendix \ref%
{Notations}. Beside the masses $m_{1}$, the number of independent variables
characterizing the total and orbital angular momenta ($\mathbf{J}$ and $%
\mathbf{L}$, respectively) and spins $\mathbf{S}_{\mathbf{i}}$ was shown to
be 6, chosen either as

(a) 3 angles (spanned by the Newtonian orbital angular momentum $\mathbf{%
L_{N}}$ with the total angular momentum $\mathbf{J}$ and with the spins $%
\mathbf{S}_{\mathbf{i}}$, denoted as $\alpha $ and $\kappa _{i}$,
respectively) and 3 scales (the normalized magnitudes of the spins $\chi
_{i}\equiv \left( c/G\right) \left( S_{i}/m_{i}^{2}\right) $ and the
magnitude of the total angular momentum $J$), or equivalently as

(b) 5 angles and a scale. In this case the dimensionless spin magnitudes $%
\chi _{i}$ could be replaced by the azimuthal angles $\psi _{i}$ of the
spins, measured in the plane of motion from a suitably defined node line $%
\mathbf{\hat{l}}$ (the intersection of the planes perpendicular to the total
orbital momentum $\mathbf{J\equiv L}+\mathbf{S}_{\mathbf{1}}+\mathbf{S}_{%
\mathbf{2}}$ and to the Newtonian orbital angular momentum $\mathbf{L_{N}}$%
). The relation between the two sets of variables is given by Eqs. (46)-(47)
of \cite{IndepVar}.

In the present paper we discuss the \textit{conservative dynamics} of these
two sets of independent variables. All lengths involved ($J$ and $\chi _{i}$%
) are constants of motion, as $\mathbf{J}$ is conserved to 2PN accuracy \cite%
{KWW} and the spins undergo a precessional motion \cite{BOC}-\cite{BOC2}.
Therefore our goal reduces to the study of the dynamics of the \textit{%
angular variables}. In the process we also derive the evolutions of the
parameters ($a_{r},e_{r}$) of the osculating ellipse; of the spin relative
angle $\gamma $ span by $\mathbf{\hat{S}}_{\mathbf{1}}$ and $\mathbf{\hat{S}}%
_{\mathbf{2}}$; and of the periastron, given by the Laplace-Runge-Lenz
vector $\mathbf{A}_{\mathbf{N}}$.

We start with a discussion of the evolutions under a generic perturbing
force in Sec. \ref{genForce}. First we monitor how the Keplerian dynamical
constants evolve. This allows us to determine both the evolution of ($%
a_{r},e_{r}$) and of the vectors $\mathbf{\hat{L}_{N}}$, $\mathbf{\hat{A}}_{%
\mathbf{N}}$, and ascending node vector $\mathbf{\hat{l}}$. We also
determine here how the evolution of the true anomaly parameter $\chi _{p}$
(measured from $\mathbf{\hat{A}}_{\mathbf{N}}$ to the actual location $%
\mathbf{\hat{r}}$ of the reduced mass particle) is modified by the
perturbing force. The specific perturbing force components generated by PN,
2PN, SO, SS and QM effects are listed in Appendix \ref{Decomp}, together
with the components of the spin precession angular velocity.

Employing these results, also the spin evolution equations discussed in
detail in \cite{IndepVar}, we are able to derive in Sec. \ref{EulerAngEvol}
the evolution of $\alpha $. Equation (14) and (15) of \cite{IndepVar} show
that once the evolution of $\chi _{p}$ and $\alpha $ are established, the
evolution of the angle $\psi _{p}$ measured from $\mathbf{\hat{l}}$ to $%
\mathbf{\hat{A}}_{\mathbf{N}}$, and of the angle $-\phi _{n}$ measured from $%
\mathbf{\hat{l}}$ to an arbitrary inertial axis $\mathbf{\hat{x}\perp J}$
(see Fig 1 of \cite{IndepVar}) also follow, which complete the
characterization of the evolution of the Euler angles. Then, in Sec. \ref%
{SpinAngEvol} we derive the evolutions of $\kappa _{i}$, $\gamma $, and $%
\psi _{i}$. With this we fulfill the task of characterizing the evolution of
the variables composing the independent sets (a) and (b).

We discuss special spin configurations in Sec. \ref{Special} and present our
Concluding Remarks in Sec. \ref{CoRe}.

\textit{Notations and conventions. }The gravitational constant $G$ and speed
of light $c$ are kept in all expressions. For any vector $\mathbf{V}$ we
denote its magnitude by $V$ and its direction by $\mathbf{\hat{V}}$.

The reduced mass is $\mu \equiv m_{1}m_{2}/m$. We assume that $m_{1}\geq
m_{2}$, thus the mass ratio $\nu \equiv m_{2}/m_{1}\leq 1$ and the symmetric
mass ratio $\eta \equiv \mu /m=\nu /\left( 1+\nu \right) ^{2}\in \left[
0,0.25\right] $.

The mass quadrupole moment originates entirely from rotation, being
therefore characterized by a single quadrupole-moment scalar $Q_{i}=-\left(
G^{2}/c^{4}\right) w\chi _{i}^{2}m_{i}^{3}$, with the parameter $w\in \left(
4,~8\right) $ for neutron stars, depending on their equation of state,
stiffer equations of state giving larger values of $w$ \cite{Poisson}, \cite%
{Larakkers Poisson 1997} and $w=1$ for rotating black holes \cite{Thorne
1980}. The negative sign arises because the rotating compact object is
centrifugally flattened, becoming an oblate spheroid.

The inertial system $\mathcal{K}_{i}$ has the arbitrary inertial $x$ axis $%
\mathbf{\hat{x}}$ and $\mathbf{\hat{J}}$ as its $z$ axis. We also define the
noninertial systems $\mathcal{K}_{L}$ and $\mathcal{K}_{A}$ with $\mathbf{%
\hat{L}}_{\mathbf{N}}$ as the common $z$ axis, the $x$ axes being $\mathbf{%
\hat{l}}$ and $\mathbf{\hat{A}}_{\mathbf{N}}$, respectively. Then the $y$
axes are $\mathbf{\hat{m}=\hat{L}}_{\mathbf{N}}\times \mathbf{\hat{l}}$ for $%
\mathcal{K}_{L}$ and $\mathbf{\hat{Q}}_{\mathbf{N}}=\mathbf{\hat{L}}_{%
\mathbf{N}}\times \mathbf{\hat{A}}_{\mathbf{N}}$ for $\mathcal{K}_{A}$.

\section{Evolutions in terms of a generic perturbing force \label{genForce}}

Although there is no notion of gravitational force within general
relativity, in the PN regime the motion of a compact binary can be regarded
as a perturbed Keplerian motion, with perturbations coming from the
difference in the predictions of general relativity with respect to
Newtonian gravity. Therefore one can adopt the terminology of celestial
mechanics, regarding the modifications induced by general relativity as
perturbing forces.

Any perturbed Keplerian motion is characterized by an acceleration%
\begin{equation}
\mathbf{a}=-\frac{Gm}{r^{2}}\mathbf{\hat{r}}+\Delta \mathbf{a~.}
\end{equation}%
We find convenient to express $\Delta \mathbf{a}$ in the basis $\mathcal{K}%
_{A}$ with basis vectors $\left\{ \mathbf{f}_{\left( \mathbf{i}\right)
}\right\} =(\mathbf{\hat{A}}_{\mathbf{N}},\ \mathbf{\hat{Q}}_{\mathbf{N}},~%
\mathbf{\hat{L}}_{\mathbf{N}})$ as%
\begin{equation}
\Delta \mathbf{a=}\sum_{i=1}^{3}a_{i}\mathbf{f}_{\left( \mathbf{i}\right) }\
.  \label{pert_force}
\end{equation}

\subsection{Keplerian dynamical constants \label{const_evol}}

Starting from the definitions of the Keplerian constants of motion $%
E_{N}\equiv \mu v^{2}/2-Gm\mu /r$, $\mathbf{L}_{\mathbf{N}}\equiv \mu 
\mathbf{r}\times \mathbf{v}$, and $\mathbf{A}_{\mathbf{N}}\equiv \mathbf{v}%
\times \mathbf{L}_{\mathbf{N}}-Gm\mu \mathbf{\hat{r}}$, it is
straightforward to show that%
\begin{equation}
\dot{E}_{N}=\mu \mathbf{v}\cdot \Delta \mathbf{a\ ,}  \label{EdotN}
\end{equation}%
\begin{equation}
\mathbf{\dot{L}_{N}}=\mu \mathbf{r}\times \Delta \mathbf{a\ ,}  \label{LdotN}
\end{equation}%
\begin{eqnarray}
\mathbf{\dot{A}}_{\mathbf{N}} &\mathbf{=}&\Delta \mathbf{a}\times \mathbf{%
L_{N}+v\times \dot{L}_{N}}  \notag \\
&=&\mu \left[ 2\left( \mathbf{v\cdot }\Delta \mathbf{a}\right) \mathbf{r-}%
\left( \mathbf{r\cdot }\Delta \mathbf{a}\right) \mathbf{v-}\left( \mathbf{%
r\cdot v}\right) \Delta \mathbf{a}\right] \mathbf{\ .}  \label{AdotN}
\end{eqnarray}%
By employing the decomposition of $\mathbf{r}$ and $\mathbf{v}$ in the basis 
$\mathcal{K}_{A}$, given by Eqs. (\ref{rKA})-(\ref{vKA}), also the
decomposition of the perturbing force acting on the unit mass (\ref%
{pert_force}), and finally the generic formula for the time-derivative of
any vector $\mathbf{V}$, 
\begin{equation}
\mathbf{\dot{V}=}\dot{V}\mathbf{\hat{V}+}V\frac{d}{dt}\mathbf{\hat{V}\ ,}
\label{timederiv}
\end{equation}%
we obtain for the magnitudes%
\begin{eqnarray}
\dot{E}_{N} &=&-a_{1}\frac{Gm\mu ^{2}}{L_{N}}\sin \chi _{p}+a_{2}\frac{\mu
\left( A_{N}+Gm\mu \cos \chi _{p}\right) }{L_{N}}\mathbf{\ ,}  \notag \\
\dot{L}_{N} &=&\left( a_{2}\cos \chi _{p}-a_{1}\sin \chi _{p}\right) \mu r%
\mathbf{\ ,}  \notag \\
\dot{A}_{N} &=&a_{2}L_{N}+\left( a_{2}\cos \chi _{p}-a_{1}\sin \chi
_{p}\right) \frac{\mu r\left( A_{N}+Gm\mu \cos \chi _{p}\right) }{L_{N}}%
\mathbf{\ .}  \label{dynconstevol}
\end{eqnarray}%
and for the directions%
\begin{eqnarray}
\frac{d}{dt}\mathbf{\hat{L}}_{\mathbf{N}} &=&a_{3}\frac{\mu }{L_{N}}r\left(
\sin \chi _{p}\mathbf{\hat{A}}_{\mathbf{N}}-\cos \chi _{p}\mathbf{\hat{Q}}_{%
\mathbf{N}}\right) ~,  \notag \\
\frac{d}{dt}\mathbf{\hat{A}}_{\mathbf{N}} &=&\left[ -a_{1}\frac{L_{N}}{A_{N}}%
+\frac{Gm\mu ^{2}}{L_{N}A_{N}}r\sin \chi _{p}\left( a_{2}\cos \chi
_{p}-a_{1}\sin \chi _{p}\right) \right] \mathbf{\hat{Q}}_{\mathbf{N}}\mathbf{%
-}a_{3}\frac{\mu r}{L_{N}}\sin \chi _{p}\mathbf{\hat{L}}_{\mathbf{N}}~,
\label{dyndirevol}
\end{eqnarray}%
where $r$ is given in terms of the true anomaly parameter $\chi _{p}$ by the
standard formula (\ref{truer}).

\subsection{Radial semimajor axis $a_{r}$ and radial eccentricity $e_{r}$}

We note that the constraint $A_{N}^{2}=\left( Gm\mu \right)
^{2}+2EL_{N}^{2}/\mu $ is preserved by the evolutions (\ref{dynconstevol}),
therefore only two of these equations are independent. From them we can also
derive evolution equations for the parameter $p_{r}=L_{N}^{2}/Gm\mu ^{2}$
and eccentricity $e_{r}=A_{N}/Gm\mu $ of the conic orbit. For bounded orbits
we could introduce the semimajor axis $a_{r}=p_{r}/\left( 1-e_{r}^{2}\right)
=L_{N}^{2}/Gm\mu ^{2}\left( 1-e_{r}^{2}\right) =-Gm\mu /2E_{N}$ of the
osculating ellipse instead, and derive evolution equations for the pair ($%
a_{r},$ $e_{r}$). In this way we obtain two \textit{Lagrange planetary
equations}:%
\begin{eqnarray}
\dot{a}_{r} &=&\frac{2a_{r}^{3/2}}{\left[ Gm\left( 1-e_{r}^{2}\right) \right]
^{1/2}}\left[ -a_{1}\sin \chi _{p}+a_{2}\left( e_{r}+\cos \chi _{p}\right) %
\right] ~,  \label{arevol} \\
\dot{e}_{r} &=&\left[ \frac{a_{r}\left( 1-e_{r}^{2}\right) }{Gm}\right]
^{1/2}\frac{a_{2}\left( 1+2e_{r}\cos \chi _{p}+\cos ^{2}\chi _{p}\right)
-a_{1}\left( e_{r}+\cos \chi _{p}\right) \sin \chi _{p}}{\left( 1+e_{r}\cos
\chi _{p}\right) }~.  \label{erevol}
\end{eqnarray}%
Here we have employed the true anomaly parametrization (\ref{truer}) written
in terms of osculating ellipse orbital elements

\begin{equation}
r=\frac{a_{r}\left( 1-e_{r}^{2}\right) }{1+e_{r}\cos \chi _{p}}~.
\label{truerosc}
\end{equation}

\subsection{The noninertial system $\mathcal{K}_{A}$}

Rewriting Eqs. (\ref{dyndirevol}) in the form of precession equations by
inserting $\mathbf{\hat{A}}_{\mathbf{N}}=\mathbf{\hat{Q}}_{\mathbf{N}}\times 
\mathbf{\hat{L}}_{\mathbf{N}}$, $\mathbf{\hat{Q}}_{\mathbf{N}}=-\mathbf{\hat{%
A}}_{\mathbf{N}}\times \mathbf{\hat{L}}_{\mathbf{N}}$ in the first
expression and $\mathbf{\hat{Q}}_{\mathbf{N}}=\mathbf{\hat{L}}_{\mathbf{N}%
}\times \mathbf{\hat{A}}_{\mathbf{N}}$, $\mathbf{\hat{L}}_{\mathbf{N}}=-%
\mathbf{\hat{Q}}_{\mathbf{N}}\times \mathbf{\hat{A}}_{\mathbf{N}}$ in the
second; also computing the time derivative of $\mathbf{\hat{Q}}_{\mathbf{N}}$
from its definition gives 
\begin{equation}
\mathbf{\dot{f}}_{(\mathbf{i})}=\mathbf{\Omega }_{A}\times \mathbf{f}_{(%
\mathbf{i})}~,  \label{KAdirevol}
\end{equation}%
with the angular velocity vector%
\begin{equation}
\mathbf{\Omega }_{A}=a_{3}\frac{\mu r\cos \chi _{p}}{L_{N}}\mathbf{\hat{A}}_{%
\mathbf{N}}+a_{3}\frac{\mu r\sin \chi _{p}}{L_{N}}\mathbf{\hat{Q}}_{\mathbf{N%
}}-\left[ a_{1}\frac{L_{N}}{A_{N}}+\left( a_{1}\sin \chi _{p}\!-\!a_{2}\cos
\chi _{p}\right) \frac{Gm\mu ^{2}r\sin \chi _{p}}{L_{N}A_{N}}\right] \mathbf{%
\hat{L}}_{\mathbf{N}}~.  \label{OmegaAVec}
\end{equation}%
With this we have established the time evolution of the noninertial basis $%
\mathcal{K}_{A}$.

The PN\ order of $\mathbf{\Omega }_{A}$ is $\mathcal{O}\left( \mathbf{\Omega 
}_{A}\right) =\varepsilon ^{-1/2}\mathcal{O}\left( a_{i}/c\right) $.
Employing the contributions to $a_{i}$ from Appendix \ref{Decomp} and Eq.
(58) of \cite{IndepVar} one finds 
\begin{eqnarray}
\mathcal{O}\left( \mathbf{\Omega }_{A}^{PN}\right) &=&\mathcal{O}\left(
\varepsilon \right) \mathcal{O}\left( 1,\eta \right) \mathcal{O}(T^{-1})~, 
\notag \\
\mathcal{O}\left( \mathbf{\Omega }_{A}^{2PN}\right) &=&\mathcal{O}\left(
\varepsilon ^{2}\right) \mathcal{O}\left( 1,\eta ,\eta ^{2}\right) \mathcal{O%
}(T^{-1})~,  \notag \\
\mathcal{O}\left( \mathbf{\Omega }_{A}^{SO}\right) &=&\mathcal{O}\left(
\varepsilon ^{3/2}\right) \left[ \sum_{k=1}^{2}\mathcal{O}\left( 1,\nu
^{2k-3}\right) \chi _{k}\right] \mathcal{O}(T^{-1})~,  \notag \\
\mathcal{O}\left( \mathbf{\Omega }_{A}^{SS}\right) &=&\mathcal{O}\left(
\varepsilon ^{2}\right) \mathcal{O}\left( \eta \right) \chi _{1}\chi _{2}%
\mathcal{O}(T^{-1})~,  \notag \\
\mathcal{O}\left( \mathbf{\Omega }_{A}^{QM}\right) &=&\mathcal{O}\left(
\varepsilon ^{2}\right) \mathcal{O}\left( \eta \right) \left[ \sum_{k=1}^{2}%
\mathcal{O}\left( \nu ^{2k-3}\right) w_{k}\chi _{k}^{2}\right] \mathcal{O}%
(T^{-1})~,  \label{OmegaAestimates}
\end{eqnarray}%
with $T$ being the radial period, defined as twice the time elapsed between
consecutive $\dot{r}=0$ configurations.

A couple of immediate remarks are in order:

(1) If $a_{3}=0$ (no perturbing force is pointing outside the plane of
motion), $\mathbf{\hat{L}}_{\mathbf{N}}$ (the plane of motion) is conserved,
while both $\mathbf{\hat{A}}_{\mathbf{N}}$ and $\mathbf{\hat{Q}}_{\mathbf{N}%
} $ undergo a precessional motion about $\mathbf{\hat{L}}_{\mathbf{N}}$ (in
the conserved plane of motion).

(2) If $a_{1}=a_{2}=0$ (the perturbing force is perpendicular to the plane
of motion), then $\mathbf{\hat{A}}_{\mathbf{N}}$ undergoes a precessional
motion about $\mathbf{\hat{Q}}_{\mathbf{N}}$ and vice-versa, while $\mathbf{%
\hat{L}}_{\mathbf{N}}$ precesses about $\mathbf{r}$.

\subsection{True anomaly $\protect\chi _{p}$}

As the basis$\left\{ \mathbf{f}_{(\mathbf{i})}\right\} $ is comoving with
the plane of motion and the periastron, the position vector $\mathbf{r}=x^{i}%
\mathbf{f}_{(\mathbf{i})}$ [with $x^{i}$ given by Eq. (\ref{rKA})] changes
according to $\mathbf{v}=\dot{x}^{i}\mathbf{f}_{(\mathbf{i})}+x^{i}\mathbf{%
\dot{f}}_{(\mathbf{i})}=\dot{x}^{i}\mathbf{f}_{(\mathbf{i})}+x^{i}\mathbf{%
\Omega }_{A}\times \mathbf{f}_{(\mathbf{i})}$. A straightforward
computation, employing 
\begin{equation}
\dot{x}^{1}=\dot{r}\cos \chi _{p}-r\dot{\chi}_{p}\sin \chi _{p}~,\qquad \dot{%
x}^{2}=\dot{r}\sin \chi _{p}+r\dot{\chi}_{p}\cos \chi _{p}~,\qquad \dot{x}%
^{3}=0~,
\end{equation}%
then leads to%
\begin{equation}
\mathbf{L}_{\mathbf{N}}=\mu r^{2}\left[ \dot{\chi}_{p}+\left( \mathbf{\Omega 
}_{A}\cdot \mathbf{\hat{L}}_{\mathbf{N}}\right) \right] \mathbf{\hat{L}}_{%
\mathbf{N}}~.  \label{LNpert}
\end{equation}%
From here 
\begin{equation}
\dot{\chi}_{p}+\left( \mathbf{\Omega }_{A}\cdot \mathbf{\hat{L}}_{\mathbf{N}%
}\right) =\frac{L_{N}}{\mu r^{2}}~,  \label{chipdot}
\end{equation}%
Therefore the deviation from the Newtonian expression is due to the
component of $\mathbf{\Omega }_{A}$ along $\mathbf{\hat{L}}_{\mathbf{N}}$.
The importance of Eq. (\ref{chipdot}) lies in allowing to pass from time
derivatives to derivatives with respect to $\chi _{p}$ in the evolution
equations (\ref{dynconstevol}), (\ref{arevol})-(\ref{erevol}), which then
become ordinary differential equations.

It is also immediate to derive $v^{2}$ and calculate $E_{N}\,$\ as 
\begin{equation}
E_{N}=\frac{\mu \left( \dot{r}^{2}+r^{2}\dot{\chi}_{p}^{2}\right) }{2}-\frac{%
Gm\mu }{r}+\mu r^{2}\dot{\chi}_{p}~\left( \mathbf{\Omega }_{A}\cdot \mathbf{%
\hat{L}}_{\mathbf{N}}\right) +\frac{\mu r^{2}}{2}\left( \mathbf{\Omega }%
_{A}\cdot \mathbf{\hat{L}}_{\mathbf{N}}\right) ^{2}{\ .}  \label{ENpert1}
\end{equation}%
By inserting Eq. (\ref{chipdot}), we obtain the radial equation 
\begin{equation}
\dot{r}^{2}=\frac{2E_{N}}{\mu }+\frac{2Gm}{r}-\frac{L_{N}^{2}}{\mu ^{2}r^{2}}%
~{.}  \label{rdot2pert2}
\end{equation}%
Remarkably, all terms arising from the precession of the basis vectors
cancelled out and we formally recovered the radial equation for the
Keplerian motion. This is not surprising, as the dynamical quantities $E_{N}$%
, $\mathbf{L}_{\mathbf{N}}$ refer to the osculating \textit{ellipse}.

\subsection{Ascending node $\mathbf{\hat{l}}$}

The basis vectors of $\mathcal{K}_{L}$ are related to the basis vectors of $%
\mathcal{K}_{A}$ by a rotation in the $x$-$y$ plane with angle $-\psi _{p}$,
thus 
\begin{eqnarray}
\mathbf{\hat{l}} &=&\cos \psi _{p}\mathbf{\hat{A}}_{\mathbf{N}}-\sin \psi
_{p}\mathbf{\hat{Q}}_{\mathbf{N}}~,  \label{l} \\
\mathbf{\hat{m}} &=&\sin \psi _{p}\mathbf{\hat{A}}_{\mathbf{N}}+\cos \psi
_{p}\mathbf{\hat{Q}}_{\mathbf{N}}~.  \label{m}
\end{eqnarray}%
The time derivative of the direction of the ascending node is therefore
found as%
\begin{eqnarray}
\frac{d}{dt}\mathbf{\hat{l}} &=&-\dot{\psi}_{p}\left( \sin \psi _{p}\mathbf{%
\hat{A}}_{\mathbf{N}}+\cos \psi _{p}\mathbf{\hat{Q}}_{\mathbf{N}}\right)
+\cos \psi _{p}\frac{d}{dt}\mathbf{\hat{A}}_{\mathbf{N}}-\sin \psi _{p}\frac{%
d}{dt}\mathbf{\hat{Q}}_{\mathbf{N}}  \notag \\
&=&\cos \psi _{p}\left( \mathbf{\Omega }_{A}-\dot{\psi}_{p}\mathbf{\hat{L}}_{%
\mathbf{N}}\right) \times \mathbf{\hat{A}}_{\mathbf{N}}-\sin \psi _{p}\left( 
\mathbf{\Omega }_{A}-\dot{\psi}_{p}\mathbf{\hat{L}}_{\mathbf{N}}\right)
\times \mathbf{\hat{Q}}_{\mathbf{N}}  \notag \\
&=&\left( \mathbf{\Omega }_{A}-\dot{\psi}_{p}\mathbf{\hat{L}}_{\mathbf{N}%
}\right) \times \mathbf{\hat{l}}~.  \label{ldirevol}
\end{eqnarray}%
Similarly we can derive the evolution of $\mathbf{\hat{m}}$ as%
\begin{eqnarray}
\frac{d}{dt}\mathbf{\hat{m}} &=&\dot{\psi}_{p}\left( \cos \psi _{p}\mathbf{%
\hat{A}}_{\mathbf{N}}-\sin \psi _{p}\mathbf{\hat{Q}}_{\mathbf{N}}\right)
+\sin \psi _{p}\frac{d}{dt}\mathbf{\hat{A}}_{\mathbf{N}}+\cos \psi _{p}\frac{%
d}{dt}\mathbf{\hat{Q}}_{\mathbf{N}}  \notag \\
&=&\cos \psi _{p}\left( \mathbf{\Omega }_{A}-\dot{\psi}_{p}\mathbf{\hat{L}}_{%
\mathbf{N}}\right) \times \mathbf{\hat{Q}}_{\mathbf{N}}+\sin \psi _{p}\left( 
\mathbf{\Omega }_{A}-\dot{\psi}_{p}\mathbf{\hat{L}}_{\mathbf{N}}\right)
\times \mathbf{\hat{A}}_{\mathbf{N}}  \notag \\
&=&\left( \mathbf{\Omega }_{A}-\dot{\psi}_{p}\mathbf{\hat{L}}_{\mathbf{N}%
}\right) \times \mathbf{\hat{m}~.}  \label{mdirevol}
\end{eqnarray}%
As it was to be expected, the unit vectors $\mathbf{\hat{l}}$ and $\mathbf{%
\hat{m}}$ undergo a precession characterized by the angular velocity vector 
\begin{equation}
\mathbf{\Omega }_{L}=\mathbf{\Omega }_{A}-\dot{\psi}_{p}\mathbf{\hat{L}}_{%
\mathbf{N}}~.  \label{OmegaLVec0}
\end{equation}

\section{Euler angle evolutions \label{EulerAngEvol}}

Now we have all necessary elements for deriving the evolution of the angles
which enter the set of independent variables. First we remark, that the time
derivative of the definition of the argument of the periastron $\psi
_{p}=\arccos \left( \mathbf{\hat{l}}\cdot \mathbf{\hat{A}}_{\mathbf{N}%
}\right) $, by employing Eqs. (\ref{KAdirevol}) and (\ref{ldirevol}) gives
an identity.

\subsection{Inclination $\protect\alpha $}

From the definition of the inclination $\alpha =\arccos \left( \mathbf{\hat{J%
}\cdot \hat{L}}_{\mathbf{N}}\right) $, employing the constancy of $\mathbf{J}
$ up to 2PN \cite{KWW} and the derived precession equation for $\mathbf{\hat{%
L}}_{\mathbf{N}}$ we find%
\begin{equation}
-\sin \alpha ~\dot{\alpha}=\mathbf{\hat{J}\cdot }\frac{d}{dt}\mathbf{\hat{L}}%
_{\mathbf{N}}=\mathbf{\hat{J}\cdot }\left( \mathbf{\Omega }_{A}\times 
\mathbf{\hat{L}}_{\mathbf{N}}\right) =~\mathbf{\Omega }_{A}\mathbf{\cdot }%
\left( \mathbf{\hat{L}}_{\mathbf{N}}\times \mathbf{\hat{J}}\right) =-\sin
\alpha ~\mathbf{\Omega }_{A}\mathbf{\cdot \hat{l}~,}
\end{equation}%
thus 
\begin{equation}
\dot{\alpha}=a_{3}\frac{\mu r\cos \left( \psi _{p}+\chi _{p}\right) }{L_{N}}%
~.  \label{alphadot}
\end{equation}

\subsection{Longitude of the ascending node $-\protect\phi _{n}$}

By employing Eq. (14) of \cite{IndepVar}\ we find the evolution of the
azimuthal angle $-\phi _{n}$ of the ascending node $\mathbf{\hat{l}}$ as\ \ 
\begin{equation}
\dot{\phi}_{n}=-a_{3}\frac{\mu r\sin \left( \psi _{p}+\chi _{p}\right) }{%
L_{N}\sin \alpha }~.  \label{phindot}
\end{equation}%
Quite naturally, both the orbital inclination and the ascending node can be
changed only by a force perpendicular to the orbit.

\subsection{Argument of the periastron $\protect\psi _{p}$}

From Eq. (15) of \cite{IndepVar}\ and Eq. (\ref{phindot}) the evolution of $%
\psi _{p}+\chi _{p}$ emerges as%
\begin{equation}
\dot{\psi}_{p}+\dot{\chi}_{p}=\frac{L_{N}}{{\mu r}^{2}}-a_{3}\frac{\mu r\sin
\left( \psi _{p}+\chi _{p}\right) }{L_{N}\tan \alpha }~.  \label{psidot}
\end{equation}%
Again, only the perturbing force component along $\mathbf{\hat{L}}_{\mathbf{N%
}}$ contributes. Combining Eqs. (\ref{psidot}) and (\ref{chipdot}) leads to
the evolution equation of the third Euler angle.%
\begin{equation}
\mathbf{\Omega }_{A}\cdot \mathbf{\hat{L}_{\mathbf{N}}}-\dot{\psi}_{p}=a_{3}%
\frac{\mu r\sin \left( \psi _{p}+\chi _{p}\right) }{L_{N}\tan \alpha }~.
\label{psipdot0}
\end{equation}%
The left-hand side is $\mathbf{\Omega }_{L}\cdot \mathbf{\hat{L}_{\mathbf{N}}%
}$, such that the unit vectors $\mathbf{\hat{l}}$ and $\mathbf{\hat{m}}$
undergo a precession characterized by the angular velocity vector 
\begin{equation}
\mathbf{\Omega }_{L}=a_{3}\frac{\mu r}{L_{N}}\left[ \cos \chi _{p}\mathbf{%
\hat{A}}_{\mathbf{N}}+\sin \chi _{p}\mathbf{\hat{Q}}_{\mathbf{N}}+\frac{\sin
\left( \psi _{p}+\chi _{p}\right) }{\tan \alpha }\mathbf{\hat{L}}_{\mathbf{N}%
}\right] ~.  \label{OmegaLVec}
\end{equation}%
The first two terms of the bracket combine to $\mathbf{\hat{r}}$. If there
is no perturbing force perpendicular to the orbit, $\mathbf{\hat{l}}$ and $%
\mathbf{\hat{m}}$ stay unchanged.

The evolution of $\psi _{p}$ in detail reads 
\begin{equation}
\dot{\psi}_{p}=-a_{1}\frac{L_{N}}{A_{N}}-\left( a_{1}\sin \chi
_{p}\!-\!a_{2}\cos \chi _{p}\right) \frac{Gm\mu ^{2}r\sin \chi _{p}}{%
L_{N}A_{N}}-a_{3}\frac{\mu r\sin \left( \psi _{p}+\chi _{p}\right) }{%
L_{N}\tan \alpha }~.  \label{psipdot}
\end{equation}

Equations (\ref{alphadot}), (\ref{phindot}) and (\ref{psipdot}) are Lagrange
planetary equations for the angular orbital elements. With the use of Eq. (%
\ref{chipdot}), by passing from time derivatives to derivatives with respect
to $\chi _{p}$, these become ordinary differential equations. During the
inspiral the perturbing force components $a_{i}$ arise as a combination of
relativistic (PN and 2PN), SO, SS and QM contributions, and are given in
Appendix \ref{Decomp}.

\section{Spin angle evolutions \label{SpinAngEvol}}

\subsection{Spin polar angles $\protect\kappa _{i}$\label{KappaIEvol}}

The spin polar angles~$\kappa _{i}=\arccos \left( \mathbf{\hat{S}}_{\mathbf{i%
}}\cdot \mathbf{\hat{L}}_{\mathbf{N}}\right) $ evolve due to the spin
precessions (see Appendix \ref{Decomp})\textbf{\ }and the evolution of $%
\mathbf{\hat{L}}_{\mathbf{N}}$, as 
\begin{equation}
-\sin \kappa _{i}~\dot{\kappa}_{i}=\left( \mathbf{\Omega }_{A}\times \mathbf{%
\hat{L}_{\mathbf{N}}}\right) \mathbf{\cdot \hat{S}_{i}+\hat{L}_{N}\cdot }%
\left( \mathbf{\Omega }_{\mathbf{i}}\times \mathbf{\hat{S}_{i}}\right)
=\left( \mathbf{\Omega }_{A}-\mathbf{\Omega }_{\mathbf{i}}\right) \cdot
\left( \mathbf{\hat{L}_{\mathbf{N}}\times \hat{S}_{i}}\right) ~.
\end{equation}%
In order to proceed, we need the expression (\ref{SiKA}) of the spin, such
that 
\begin{equation}
\mathbf{\hat{L}_{\mathbf{N}}\times \hat{S}_{i}}=\sin \kappa _{i}\left[ \sin
\left( \psi _{p}-\psi _{i}\right) \mathbf{\hat{A}}_{\mathbf{N}}+\cos \left(
\psi _{p}-\psi _{i}\right) \mathbf{\hat{Q}}_{\mathbf{N}}\right] ~,
\label{LNcrossSi}
\end{equation}%
and we find%
\begin{equation}
~\dot{\kappa}_{i}=\left( \mathbf{\Omega }_{\mathbf{i}}\cdot \mathbf{\hat{A}}%
_{\mathbf{N}}\right) \sin \left( \psi _{p}-\psi _{i}\right) +\left( \mathbf{%
\Omega }_{\mathbf{i}}\cdot \mathbf{\hat{Q}}_{\mathbf{N}}\right) \cos \left(
\psi _{p}-\psi _{i}\right) -a_{3}\frac{\mu r}{L_{N}}\sin \left( \psi
_{p}+\chi _{p}-\psi _{i}\right) ~.  \label{kappaidot}
\end{equation}%
The relative orientation of spins with respect to the orbital angular
momentum is unchanged only if the perturbing force lies in the plane of
motion ($a_{3}=0$) and if the spin precession axis is along $\mathbf{\hat{L}%
_{\mathbf{N}}}$. The latter condition is obeyed by the SO precession, but
not by its SS and QM corrections (except for perfect perpendicularity of the
spins to the orbital plane, when also $a_{3}=0$ holds, see Appendix \ref%
{Decomp}, thus $\dot{\kappa}_{i}=0$). Starting from this and the remark $%
a_{3}\propto \mathcal{O}\left( \varepsilon ^{3/2}\right) $, also the
estimates (\ref{LS1S2estimates}) we find 
\begin{equation}
\mathcal{O}\left( \dot{\kappa}_{i}\right) =\mathcal{O}\left( \varepsilon
^{3/2}\right) \mathcal{O}\left( \eta \right) \left[ w_{i}\chi _{i}+\mathcal{O%
}\left( \nu ^{2i-3}\right) \chi _{j}~\right] \mathcal{O}(T^{-1})~.~
\label{kappadotestimate}
\end{equation}

\subsection{Relative spin angle $\protect\gamma $\label{GammaEvol}}

For this we take the derivative of its definition $\gamma =\arccos \left( 
\mathbf{\hat{S}_{1}\cdot \hat{S}_{2}}\right) $ and obtain 
\begin{equation}
-\sin \gamma ~\dot{\gamma}=\left( \mathbf{\Omega }_{\mathbf{1}}-\mathbf{%
\Omega }_{\mathbf{2}}\right) \cdot \left( \mathbf{\hat{S}_{1}\times \hat{S}%
_{2}}\right) ~.  \label{gammaconst}
\end{equation}%
If the spins are either aligned or antialigned with each other, such that $%
\mathbf{\hat{S}_{1}}\times \mathbf{\hat{S}_{2}}=0$, then $\dot{\gamma}=0$,
irrespective of the mass ratio.

Otherwise, by employing Eqs. (56) of \cite{IndepVar}\textbf{\ }and also $%
\left( \mathbf{\hat{S}_{1}}\times \mathbf{\hat{S}_{2}}\right) \cdot \mathbf{%
\hat{S}_{i}}=0$, we rewrite the condition (\ref{gammaconst}) as 
\begin{equation}
-\frac{c^{2}r^{3}}{3G}\sin \gamma ~\dot{\gamma}=\left( {\frac{\left( \nu
-\nu ^{-1}\right) }{2}}\mathbf{L_{N}}+\left\{ \mathbf{\hat{r}}\cdot \left[
\left( 1\mathbf{\mathbf{-}}w_{2}\nu ^{-1}\right) \mathbf{S}_{\mathbf{2}%
}-\left( 1-w_{1}\nu \right) \mathbf{S}_{\mathbf{1}}\right] \right\} \mathbf{%
\hat{r}}\right) \cdot \left( \mathbf{\hat{S}_{1}\times \hat{S}_{2}}\right) ~.
\label{condgamma}
\end{equation}%
Equal mass ($\nu =1$) black holes ($w_{i}=1$) trivially imply $\dot{\gamma}%
=0 $, irrespective of the orientations of the spins.

For the generic case from Eq. (\ref{SiKA}) we have 
\begin{eqnarray}
\mathbf{\hat{S}_{1}}\times \mathbf{\hat{S}_{2}} &=&\left[ \cos \kappa
_{1}\sin \kappa _{2}\sin \left( \psi _{p}-\psi _{2}\right) -\sin \kappa
_{1}\cos \kappa _{2}\sin \left( \psi _{p}-\psi _{1}\right) \right] \mathbf{%
\hat{A}}_{\mathbf{N}}  \notag \\
&&+\left[ \cos \kappa _{1}\sin \kappa _{2}\cos \left( \psi _{p}-\psi
_{2}\right) -\sin \kappa _{1}\cos \kappa _{2}\cos \left( \psi _{p}-\psi
_{1}\right) \right] \mathbf{\hat{Q}}_{\mathbf{N}}  \notag \\
&&+\sin \kappa _{1}\sin \kappa _{2}\sin \left( \psi _{2}-\psi _{1}\right) 
\mathbf{\hat{L}}_{\mathbf{N}}~,
\end{eqnarray}%
then 
\begin{eqnarray}
\left( \mathbf{\hat{S}_{1}}\times \mathbf{\hat{S}_{2}}\right) \cdot \mathbf{%
\hat{r}} &\mathbf{=}&\cos \kappa _{1}\sin \kappa _{2}\sin \left( \psi
_{p}+\chi _{p}-\psi _{2}\right) -\sin \kappa _{1}\cos \kappa _{2}\sin \left(
\psi _{p}+\chi _{p}-\psi _{1}\right) ~,  \notag \\
\left( \mathbf{\hat{S}_{1}}\times \mathbf{\hat{S}_{2}}\right) \cdot \mathbf{%
\hat{L}}_{\mathbf{N}} &=&\sin \kappa _{1}\sin \kappa _{2}\sin \left( \psi
_{2}-\psi _{1}\right) ~.
\end{eqnarray}%
Thus we can rewrite Eq. (\ref{condgamma}) in detail as%
\begin{eqnarray}
-\frac{c^{2}r^{3}}{3GL_{N}}\sin \gamma ~\dot{\gamma} &=&{\frac{\left( \nu
-\nu ^{-1}\right) }{2}}\sin \kappa _{1}\sin \kappa _{2}\sin \left( \psi
_{2}-\psi _{1}\right)  \notag \\
&&+\left[ \left( 1-w_{2}\nu ^{-1}\right) \frac{S_{2}}{{L}_{N}}\sin \kappa
_{2}\cos \left( \psi -\psi _{2}\right) -\left( 1-w_{1}\nu \right) \frac{S_{1}%
}{{L}_{N}}\sin \kappa _{1}\cos \left( \psi -\psi _{1}\right) \right]  \notag
\\
&&\times \left[ \cos \kappa _{1}\sin \kappa _{2}\sin \left( \psi -\psi
_{2}\right) -\sin \kappa _{1}\cos \kappa _{2}\sin \left( \psi -\psi
_{1}\right) \right] ~,  \label{condgamma1}
\end{eqnarray}%
where $\psi =\psi _{p}+\chi _{p}$. Again, it is manifest, that the relative
angle of the spins stays constant for equal mass black holes, irrespective
of their orientation.

Starting from the above remark, Eq. (\ref{gammaconst}) and the estimates (%
\ref{LS1S2estimates}) we find 
\begin{equation}
\mathcal{O}\left( \dot{\gamma}\right) =\mathcal{O}\left( \varepsilon \right) 
\mathcal{O}\left( \nu ^{3-2i}\right) \mathcal{O}(T^{-1})~.~
\end{equation}%
Thus the angle $\gamma $ changes faster than $\kappa _{i}$.

\subsection{Spin azimuthal angles $\protect\psi _{i}$\label{PsiIEvol}}

Eq. (30) of \cite{IndepVar}\ 
\begin{equation}
\mathbf{\hat{S}_{i}}=\sin \kappa _{i}\cos \psi _{i}\mathbf{\hat{l}}+\sin
\kappa _{i}\sin \psi _{i}\mathbf{\hat{m}+}\cos \kappa _{i}\mathbf{\hat{L}}_{%
\mathbf{N}}~
\end{equation}%
gives $\psi _{i}=\arctan \left[ \left( \mathbf{\hat{m}}\cdot \mathbf{\hat{S}%
_{i}}\right) /\left( \mathbf{\hat{l}}\cdot \mathbf{\hat{S}}_{\mathbf{i}%
}\right) \right] $ for the spin azimuthal angles, unless $\kappa _{i}=0,\pi $
(the spins are aligned or antialigned to the Newtonian orbital angular
momentum) or $\psi _{i}=\pi /2,3\pi /2$ (the projections of the spins in the
plane of motion are perpendicular to the node line).

In the generic case the spin azimuthal angles evolve according to%
\begin{equation}
\left( 1+\tan ^{2}\psi _{i}\right) \dot{\psi}_{i}\left( \mathbf{\hat{l}}%
\cdot \mathbf{\hat{S}_{i}}\right) =\left( \mathbf{\hat{m}}-\tan \psi _{i}%
\mathbf{\hat{l}}\right) \cdot \frac{d}{dt}\mathbf{\hat{S}_{i}+}\left( \frac{d%
}{dt}\mathbf{\hat{m}}-\tan \psi _{i}\frac{d}{dt}\mathbf{\hat{l}}\right) 
\mathbf{\cdot \mathbf{\hat{S}_{i}}~.}
\end{equation}%
As both $\mathbf{\hat{l}}$ and $\mathbf{\hat{m}}$ precesses about $\mathbf{%
\Omega }_{L}$, while $\mathbf{\hat{S}_{i}}$ about $\mathbf{\Omega }_{\mathbf{%
i}}$, we find%
\begin{equation}
\dot{\psi}_{i}\sin \kappa _{i}=\left( \cos \psi _{i}\mathbf{\hat{m}}-\sin
\psi _{i}\mathbf{\hat{l}}\right) \cdot \left[ \left( \mathbf{\Omega }_{%
\mathbf{i}}-\mathbf{\Omega }_{L}\right) \times \mathbf{\hat{S}_{i}}\right] 
\mathbf{~,}
\end{equation}%
or, by employing Eqs. (\ref{l})-(\ref{m}):%
\begin{equation}
\dot{\psi}_{i}\sin \kappa _{i}=\left[ \sin \left( \psi _{p}-\psi _{i}\right) 
\mathbf{\hat{A}}_{\mathbf{N}}+\cos \left( \psi _{p}-\psi _{i}\right) \mathbf{%
\hat{Q}}_{\mathbf{N}}\right] \cdot \left[ \left( \mathbf{\Omega }_{\mathbf{i}%
}-\mathbf{\Omega }_{L}\right) \times \mathbf{\hat{S}_{i}}\right] \mathbf{~,}
\label{psiidot0}
\end{equation}%
with the vector products $\mathbf{\Omega }_{\mathbf{i}}\times \mathbf{\hat{S}%
_{i}}$ and $\mathbf{\Omega }_{L}\times \mathbf{\hat{S}_{i}}$, given by Eqs. (%
\ref{OmitimesSi}) and (\ref{OmLtimesSi}), respectively. We obtain%
\begin{eqnarray}
\dot{\psi}_{i} &=&\left( \mathbf{\Omega }_{\mathbf{i}}\cdot \mathbf{\hat{L}}%
_{\mathbf{N}}\right) +\left[ \left( \mathbf{\Omega }_{\mathbf{i}}\cdot 
\mathbf{\hat{Q}}_{\mathbf{N}}\right) \sin \left( \psi _{p}-\psi _{i}\right)
-\left( \mathbf{\Omega }_{\mathbf{i}}\cdot \mathbf{\hat{A}}_{\mathbf{N}%
}\right) \cos \left( \psi _{p}-\psi _{i}\right) \right] \cot \kappa _{i} 
\notag \\
&&-a_{3}\frac{\mu r}{L_{N}}\left[ \cot \alpha \sin \left( \psi _{p}+\chi
_{p}\right) -\cot \kappa _{i}\cos \left( \chi _{p}+\psi _{p}-\psi
_{i}\right) \right] \mathbf{~.}  \label{psiidot}
\end{eqnarray}%
With this we have completed the derivation of all required evolution
equations.

Starting from Eq. (\ref{psiidot}) and the estimates (\ref{LS1S2estimates})
we find 
\begin{equation}
\mathcal{O}\left( \dot{\psi}_{i}\right) =\mathcal{O}\left( \varepsilon
\right) \mathcal{O}\left( 1,\eta \right) \mathcal{O}(T^{-1})~.~
\end{equation}%
The change in the azimuthal angle of the spins is one PN order higher than
the Keplerian orbital evolution.

\section{Special configurations\label{Special}}

As a by-product of the calculations carried on in this paper we have
recovered the known result that the plane of motion is changed only by
perturbing forces pointing outside the plane of motion, thus by the SO, SS
and QM perturbations. We have shown that the relative angle of the spins
stays constant for equal mass black holes, irrespective of their
orientation. We have also proven that unless the spins are perpendicular to
the plane of motion ($\kappa _{i}=0$), the polar spin angles will change
under these perturbations.

The nonprecessing ($\kappa _{i}=0$) and precessing (generic $\kappa _{i}$)
cases have been discussed separately in the literature (see Refs \cite%
{recoilSpinAlignedL} and \cite{recoilSpinningOrbitalPlane}, respectively) in
connection with the recoil of the final black hole \cite%
{recoilSpinningAnalytical}. From among the precessing cases the antialigned
spin configuration with the spins laying in the orbital plane has received
special attention, as numerical investigations have shown that it leads to
the highest kick velocity.

We have now the means to investigate such a configuration analytically.
First we specialize to spins laying in the orbital plane, $\kappa _{i}=\pi
/2 $. After some algebra, Eq. (\ref{kappaidot}) gives 
\begin{eqnarray}
~\dot{\kappa}_{i} &=&\frac{G^{2}m^{2}\eta }{2c^{3}r^{3}}\left(
K_{i}^{SO}+K_{i}^{SS}+K_{i}^{QM}\right) ~,  \label{Ks} \\
K_{i}^{SO} &=&-\frac{\sin \left( \psi _{p}+\chi _{p}-\psi _{i}\right) }{1+%
\frac{A_{N}}{Gm\mu }\cos \chi _{p}}\sum_{k=1}^{2}\left( 4\nu
^{2k-3}+3\right) \chi _{k}  \notag \\
&&\times \left[ 2\cos \left( \psi _{p}+\chi _{p}-\psi _{k}\right) +\frac{%
A_{N}}{Gm\mu }\left[ 2\cos \left( \psi _{p}-\psi _{k}\right) -3\sin \chi
_{p}\sin \left( \psi _{p}+\chi _{p}-\psi _{k}\right) \right] \right] ~, 
\notag \\
K_{i}^{SS} &=&\nu ^{2j-3}\chi _{j}\left[ 3\sin \left( 2\psi _{p}+2\chi
_{p}-\psi _{i}-\psi _{j}\right) +\sin \left( \psi _{j}-\psi _{i}\right) %
\right] ~,  \notag \\
K_{i}^{QM} &=&3w_{i}\chi _{i}\sin \left( 2\psi _{p}+2\chi _{p}-2\psi
_{i}\right) ~.  \notag
\end{eqnarray}%
All contributions $K_{i}^{SO}$, $K_{i}^{SS}$, $K_{i}^{QM}$ are of the same
order. In general the expression for $\dot{\kappa}_{i}$ does not vanish, not
even in the special case of equal mass ($\nu =1$), maximally spinning ($\chi
_{i}=1$) black holes ($w_{i}=1$) on circular orbit ($A_{N}=0$), when 
\begin{equation}
~\dot{\kappa}_{i}=-\frac{G^{2}m^{2}\eta }{c^{3}r^{3}}\left[ 2\sin \left(
2\psi _{p}+2\chi _{p}-2\psi _{i}\right) +2\sin \left( 2\psi _{p}+2\chi
_{p}-\psi _{i}-\psi _{j}\right) +3\sin \left( \psi _{j}-\psi _{i}\right) %
\right] ~.
\end{equation}%
Therefore in general a configuration with the spins in the plane of motion
is not preserved.

However in the special case $\psi _{j}=\psi _{i}+\pi $ and equal mass ($\nu
=1$), equal spin ($\chi _{2}=\chi _{1}$) black holes ($w_{i}=1$) we find 
\begin{eqnarray}
a_{3} &=&0~,  \notag \\
\mathbf{\Omega }_{\mathbf{i}}\cdot \mathbf{\hat{A}}_{\mathbf{N}} &=&\frac{%
G^{2}m^{2}\eta }{c^{3}r^{3}}\chi _{1}\cos \left( \psi _{p}-\psi _{i}\right)
~,  \notag \\
\mathbf{\Omega }_{\mathbf{i}}\cdot \mathbf{\hat{Q}}_{\mathbf{N}} &=&-\frac{%
G^{2}m^{2}\eta }{c^{3}r^{3}}\chi _{1}\sin \left( \psi _{p}-\psi _{i}\right) 
\mathbf{~,}  \notag \\
\mathbf{\Omega }_{\mathbf{i}}\cdot \mathbf{\hat{L}}_{\mathbf{N}} &=&{\frac{7G%
}{2c^{2}r^{3}}}J\cos \alpha ~,
\end{eqnarray}%
such that according to Eq. (\ref{kappaidot}) $\dot{\kappa}_{i}=0$.\footnote{%
The SO contribution to $\dot{\kappa}_{i}$ vanishes, while the SS and QM
contributions cancel. A glance at $K_{i}^{QM}$ given by Eqs. (\ref{Ks})
shows that without imposing the black hole condition $w_{i}=1$ the SS and QM
contributions do \textit{not} cancel, therefore the result does not hold for
equal mass, identically spinning neutron stars.}

Then one has to check, whether the condition imposed on $\psi _{i}$ is
consistent with their evolution. With $a_{3}=0$ Eq. (\ref{OmLtimesSi}) gives 
$\mathbf{\Omega }_{L}\times \mathbf{\hat{S}_{i}}=0$, while from Eq. (\ref%
{OmitimesSi}) we get%
\begin{equation}
\mathbf{\Omega }_{\mathbf{i}}\times \mathbf{\hat{S}_{i}}={\frac{7G}{%
2c^{2}r^{3}}}J\cos \alpha \left[ \sin \left( \psi _{p}-\psi _{i}\right) 
\mathbf{\hat{A}}_{\mathbf{N}}+\cos \left( \psi _{p}-\psi _{i}\right) \mathbf{%
\hat{Q}}_{\mathbf{N}}\right] ~,
\end{equation}%
such that Eq. (\ref{psiidot}) simplifies to%
\begin{equation}
\dot{\psi}_{i}={\frac{7G}{2c^{2}r^{3}}}J\cos \alpha \mathbf{~.}
\end{equation}%
As the right-hand side does not depend on the index $i$, the imposed
antialignment of the spins can be maintained over time. This is also evident
from Eq. (\ref{gammaconst}). We have also checked that the constraints
(46)-(47) of \cite{IndepVar} are trivially obeyed.

Therefore the special configuration of \textit{equal mass black holes with
equal, but antialigned spins, both laying in the plane of motion is
preserved by the conservative PN dynamics, with leading order SO, SS and QM
contributions included}. This stands as the main result of this section.

Equation (48) of \cite{IndepVar} allows us to rewrite 
\begin{equation}
\dot{\psi}_{i}={\frac{7G}{2c^{2}r^{3}}}L_{N}\left( 1+\epsilon _{PN}+\epsilon
_{2PN}\right) \mathbf{~,}
\end{equation}%
with the coefficients (given by Eqs. (39)-(40) of \cite{IndepVar}) specified
for equal mass as%
\begin{eqnarray}
\epsilon _{PN} &=&\frac{1}{8}\left( \frac{v}{c}\right) ^{2}+\frac{13}{4}%
\frac{Gm}{c^{2}r}~,  \label{epPNeqmass} \\
\epsilon _{2PN} &=&\frac{3}{128}\left( \frac{v}{c}\right) ^{4}-\frac{13}{32}%
\frac{Gm}{c^{2}r}\left( \frac{\dot{r}}{c}\right) ^{2}+\frac{63}{32}\frac{Gm}{%
c^{2}r}\left( \frac{v}{c}\right) ^{2}+\left( \frac{Gm}{c^{2}r}\right) ^{2}~.
\label{ep2PNeqmass}
\end{eqnarray}

\section{Concluding Remarks\label{CoRe}}

In this paper we have established the conservative evolution equations of
the two independent sets of variables characterizing a spinning compact
binary during its inspiral, established in \cite{IndepVar}, with leading
order SO, SS and QM contributions included.\ As the lengths $J$ and $\chi
_{i}$ are constants, this reduces to angular evolutions. The evolutions of
the variables complementing the set ($J,~\chi _{i})$, the inclination $%
\alpha $ and the spin polar angles $\kappa _{i}$ were given as Eqs. (\ref%
{alphadot}) and (\ref{kappaidot}). The evolution equations for the spin
azimuthal angles $\psi _{i}\,$\ (replacing $\chi _{i}$ as independent
variables) were given by Eq. (\ref{psiidot}). These time derivatives (and
all others computed throughout the paper) can be transformed to derivatives
with respect to $\chi _{p}$ by employing Eq. (\ref{chipdot}) in the form 
\begin{equation}
\frac{d}{dt}=\left( \frac{L_{N}}{\mu r^{2}}-\mathbf{\Omega }_{A}\cdot 
\mathbf{\hat{L}}_{\mathbf{N}}\right) \frac{d}{d\chi _{p}}~.  \label{timechi}
\end{equation}%
The true anomaly $\chi _{p}$ becomes the only independent variable by
employing the parametrization $r\left( \chi _{p}\right) $, Eqs. (\ref{truer}%
)-(\ref{truerdot}).

The system is closed by the evolution of the argument of the periastron $%
\psi _{p}$\thinspace\ given as Eq. (\ref{psipdot}), the last two Eqs. (\ref%
{dynconstevol}) giving $\dot{A}_{N}\,$\ and $\dot{L}_{N}$, the analytical
expression (\ref{a1})-(\ref{a3}) of the perturbing acceleration components
\thinspace $a_{i}$, the expressions (\ref{Omiproj1})-(\ref{Omiproj2}) of the
of the spin precessional angular velocity components $\left( \mathbf{\Omega }%
_{\mathbf{i}}\cdot \mathbf{\hat{A}}_{\mathbf{N}}\right) $ and $\left( 
\mathbf{\Omega }_{\mathbf{i}}\cdot \mathbf{\hat{Q}}_{\mathbf{N}}\right) $,
finally the vector products $\mathbf{\Omega }_{\mathbf{i}}\times \mathbf{%
\hat{S}_{i}}$ and $\mathbf{\Omega }_{L}\times \mathbf{\hat{S}_{i}}$, given
by Eqs. (\ref{OmitimesSi}) and (\ref{OmLtimesSi}), respectively.

Therefore we have derived a \textit{closed system of first order ordinary
differential equations} for the variables ($\alpha $, $\kappa _{i}$, $\psi
_{i}$, $\psi _{p}$, $A_{N}$, $L_{N}$) evolving in terms of the true anomaly $%
\chi _{p}$, ready for numerical evolution. From this set ($\alpha $, $\kappa
_{i}$, $\psi _{i}$) are independent variables characterizing the spinning
binary configuration, while ($\psi _{p}$, $A_{N}$, $L_{N}$) characterize the
orbit.

In another way of counting, replacing ($A_{N}$, $L_{N}$) and their
evolutions by the orbital elements (\thinspace $a_{r}$, $e_{r}$) and Eqs. (%
\ref{arevol})-(\ref{erevol}), respectively; also including the evolution Eq.
(\ref{phindot}) for the longitude of the ascending node $-\phi _{n}$ we have
obtained evolutions for (i) the orbital elements (\thinspace $a_{r}$, $e_{r}$%
, $\alpha $, $\psi _{p}$, $-\phi _{n}$) characterizing the perturbed
Keplerian motion and for (ii) the spin angles ($\kappa _{i}$, $\psi _{i}$)
characterizing the spin orientations with respect to this perturbed
Keplerian orbit.

As a by-product, we have proven that the relative angle of the spins stays
constant for equal mass black holes, irrespective of their orientation.

Also, unless the spins are perpendicular to the plane of motion, the polar
spin angles change under the perturbations. There is one notable exception
under this rule: the special configuration of equal mass black holes with
equal, but antialigned spins, both laying in the plane of motion is
preserved by the conservative dynamics. This is the configuration which led
to maximal recoil found in numerical simulations \cite%
{recoilSpinningOrbitalPlane}, and our investigations show that it is
conserved during the inspiral to a 2PN accuracy, with leading order
spin-orbit, spin-spin and mass quadrupole effects included.

\section*{Acknowledgements}

I am grateful to Thomas Krichbaum for references on observational spin
estimates. This work was partially supported by COST Action MP0905 "Black
Holes in a Violent Universe" and Hungarian Scientific Research Fund (OTKA)
Grant no. 69036.

\appendix

\section{Comparison of notations with related literature\label{Notations}}

In this Appendix we compare the notations established in \cite{IndepVar} and
thoroughly employed in this paper with corresponding notations in the
literature.

First we establish the correspondence of the Euler angles $\left( -\phi
_{n},\alpha ,\psi _{p}\right) $ employed in \cite{IndepVar} and standard
celestial mechanics angular orbital elements in Table \ref{table1}. The
celestial mechanics angular orbital elements $\left( \Omega ,\iota ,\omega
\right) $ are defined with respect to a reference plane and a reference
direction contained within it, both inertial. The node line is defined as
the intersection of the reference plane with the plane of motion; the angle
span by it with the reference direction is the \textit{longitude of the
ascending node} $\Omega $; the relative angle of the two planes is the 
\textit{inclination} $\iota $ and the angle span by the ascending node with
the direction of the periastron in the \textit{argument of the periastron} $%
\omega $. The Euler angles $\left( -\phi _{n},\alpha ,\psi _{p}\right) $
employed in \cite{IndepVar} are defined similarly, but with respect to the
inertial system $\mathcal{K}_{i}$ with $\mathbf{\hat{x}}$ and $\mathbf{\hat{J%
}}$ standing as the $x$ and $z$ axes [any $\mathbf{\hat{x}\perp }$ $\mathbf{%
\hat{J}}$ standing as the reference direction and the reference plane given
by $\left( \mathbf{\hat{x}},\mathbf{\hat{y}}=\mathbf{\hat{J}}\times \mathbf{%
\hat{x}}\right) $].

\begin{table}[h]
\caption{Comparison of the notations in Paper \textbf{I} and standard
celestial mechanics angular orbital elements. }
\label{table1}%
\begin{tabular}{ll|ll}
\multicolumn{2}{l}{Ref. \cite{IndepVar}} & \multicolumn{2}{l}{Celestial
mechanics} \\ \hline
Euler angles & $\left( -\phi _{n},\alpha ,\psi _{p}\right) $ & angular
orbital elements & $\left( \Omega ,\iota ,\omega \right) $ \\ 
True anomaly & $\chi _{p}$ & true anomaly & $v$ \\ 
Equation (14) of \cite{IndepVar} & $\dot{\phi}_{n}=-\dot{\alpha}\frac{\tan
\left( \psi _{p}+\chi _{p}\right) }{\sin \alpha }$ &  & $\dot{\Omega}=\dot{%
\iota}\frac{\tan \left( \omega +v\right) }{\sin \iota }$ \\ 
Equation (15) of \cite{IndepVar} & $\dot{\psi}_{p}+\dot{\chi}_{p}=\frac{L_{N}%
}{\mu r^{2}}+\dot{\phi}_{n}\cos \alpha $ &  & $\dot{\omega}+\dot{v}=\frac{%
L_{N}}{\mu r^{2}}-\dot{\Omega}\cos \iota $%
\end{tabular}%
\end{table}

The various systems of reference necessary for the description of the motion
were also discussed in Refs. \cite{FC} and \cite{BCV2}. We establish the
correspondence in Table \ref{table2}. While in these papers a quasicircular
orbit was assumed, the results of \cite{IndepVar} hold for generic orbits. A
correspondence can be established as long as $\mathbf{\hat{J}}$ can be
viewed as an inertial axis.

\begin{table}[h]
\caption{Comparison of the notations in\ Refs. \protect\cite{FC} and 
\protect\cite{BCV2} and \protect\cite{IndepVar}.}
\label{table2}%
\begin{tabular}{l|l|l}
& Reference \cite{BCV2} (based on \cite{FC}) & Reference \cite{IndepVar} \\ 
\hline
Orbit & circular & elliptical, $A_{N}\neq 0$ \\ 
Corresponding quantities & $\left( \omega ,~M\omega \right) $ & $\left( 
\frac{L_{N}}{\mu r^{2}}=\frac{c^{3}}{Gm}\varepsilon ^{3/2},~\frac{c^{3}}{G}%
\varepsilon ^{3/2}\right) $ \\ 
Corresponding equation numbers & (16), (18) & (25) and (26), (15) \\ 
Plane of motion & orthonormal base $\left( n,\lambda \right) $ & 
nonorthogonal base $\left( \frac{r}{r},\frac{v}{v}\right) $ \\ 
Orthonormal inertial source system & $\left(
e_{x}^{S},e_{y}^{S},e_{z}^{S}\right) $ & $K_{i}=\left( \hat{x},\hat{y},\hat{J%
}\right) $, if $\hat{J}\equiv e_{z}^{S}$ \\ 
Orthonormal basis in the plane of motion & $\left(
e_{1}^{S},e_{2}^{S}\right) $ & $\left( \hat{l},\hat{m}\right) $, if $\hat{J}%
\equiv e_{z}^{S}$ \\ 
Euler angles & $\left( \Phi _{S},\iota ,\alpha \right) $ & $\left( \psi
=\psi _{p}+\chi _{p},\alpha ,\frac{\pi }{2}-\phi _{n}\right) $, if $\hat{J}%
\equiv e_{z}^{S}$ \\ 
Line of sight in the $\left( x,z\right) $ plane & $\Theta =\arccos \left(
e_{z}^{S}\cdot \hat{N}\right) $ & $\Theta =\arccos \left( \hat{J}\cdot \hat{N%
}\right) ,~\hat{y}=\frac{\hat{J}\times \hat{N}}{\sin \Theta }$%
\end{tabular}%
\end{table}

Finally we compare the notations of Ref. \cite{ABFO} with the notations of 
\cite{IndepVar} in Table \ref{table3}.

\begin{table}[h]
\caption{Comparison of the notations in\ Refs. \protect\cite{ABFO} and 
\protect\cite{IndepVar}.}
\label{table3}%
\begin{tabular}{l|l|l}
& Reference \cite{ABFO} & Reference \cite{IndepVar} \\ \hline
Orbit & circular & elliptical, $A_{N}\neq 0$ \\ 
Corresponding quantities & $\left( \omega _{orb},~v^{3}=M\omega
_{orb}\right) $ & $\left( \frac{L_{N}}{\mu r^{2}}=\frac{c^{3}}{Gm}%
\varepsilon ^{3/2},~\frac{c^{3}}{G}\varepsilon ^{3/2}\right) $ \\ 
Inertial axis & $J_{0}$ & $\hat{J}$ \\ 
Inertial orthonormal basis $\perp $ $J_{0}$ (to $\hat{J}$) & $\left( \hat{x}%
,~\hat{y}\right) $ & $\left( \hat{x},~\hat{y}\right) $ \\ 
Inertial orthonormal basis $\perp \hat{L}_{N}$ (the $K_{L}$ basis) & $\left( 
\hat{x}_{L},~\hat{y}_{L}\right) $ & $\left( \hat{l},~\hat{m}\right) $ \\ 
Basis comoving with $\mu $ & $\left( n,\lambda \right) $ & $\left( \hat{r},%
\hat{L}_{N}\times \hat{r}\right) $ \\ 
Symmetric mass ratio & $\nu \in \left[ 0,0.25\right] $ & $\eta \in \left[
0,0.25\right] $ \\ 
Polar and azimuthal angles of $\hat{L}_{N}$ in $K_{i}$ & $\left( \iota
,\alpha \right) $ & $\left( \alpha ,\frac{3\pi }{2}-\phi _{n}\right) $ \\ 
Phase & $\Phi \left( t\right) $ & $\psi =\psi _{p}+\chi _{p}$%
\end{tabular}%
\end{table}

\section{Decomposition of the acceleration and spin angular velocity vectors
in the system $\mathcal{K}_{A}$ during the inspiral\label{Decomp}}

In this Appendix we give the decomposition of the accelerations and of the
precessional angular velocities of the spins in the system $\mathcal{K}_{A}$%
. The ingredients we need are Eqs. (19)-(20) of \cite{IndepVar} for the
decomposition of the position and velocity vectors:%
\begin{eqnarray}
\mathbf{\hat{r}} &=&\cos \chi _{p}\mathbf{\hat{A}}_{\mathbf{N}}+\sin \chi
_{p}\mathbf{\hat{Q}}_{\mathbf{N}}\mathbf{~,}  \label{rKA} \\
\mathbf{v} &=&\frac{Gm\mu }{L_{N}}\left[ -\sin \chi _{p}\mathbf{\hat{A}}_{%
\mathbf{N}}+\left( \cos \chi _{p}+\frac{A_{N}}{Gm\mu }\right) \mathbf{\hat{Q}%
}_{\mathbf{N}}\right] \ .  \label{vKA}
\end{eqnarray}%
In the system $\mathcal{K}_{L}$ the spin is given by Eq. (30) of \cite%
{IndepVar}. By employing Eqs. (\ref{l})-(\ref{m}) we rewrite it in the
system $\mathcal{K}_{A}$ as%
\begin{equation}
\mathbf{\hat{S}_{i}}=\sin \kappa _{i}\left[ \cos \left( \psi _{p}-\psi
_{i}\right) \mathbf{\hat{A}}_{\mathbf{N}}-\sin \left( \psi _{p}-\psi
_{i}\right) \mathbf{\hat{Q}}_{\mathbf{N}}\right] +\cos \kappa _{i}\mathbf{%
\hat{L}}_{\mathbf{N}}~.  \label{SiKA}
\end{equation}%
We also need%
\begin{eqnarray}
\mathbf{\hat{r}}\times \mathbf{\hat{S}}_{\mathbf{k}} &=&\cos \kappa
_{k}\left( \sin \chi _{p}\mathbf{\hat{A}}_{\mathbf{N}}-\cos \chi _{p}\mathbf{%
\hat{Q}}_{\mathbf{N}}\right) -\sin \kappa _{k}\sin \left( \psi _{p}+\chi
_{p}-\psi _{k}\right) \mathbf{\hat{L}}_{\mathbf{N}}~, \\
\mathbf{v\times \hat{S}}_{\mathbf{k}} &=&\frac{Gm\mu }{L_{N}}\cos \kappa _{k}%
\left[ \left( \cos \chi _{p}+\frac{A_{N}}{Gm\mu }\right) \mathbf{\hat{A}}_{%
\mathbf{N}}+\sin \chi _{p}\mathbf{\hat{Q}}_{\mathbf{N}}\right]  \notag \\
&&-\frac{Gm\mu }{L_{N}}\sin \kappa _{k}\left[ \cos \left( \psi _{p}+\chi
_{p}-\psi _{k}\right) +\frac{A_{N}}{Gm\mu }\cos \left( \psi _{p}-\psi
_{k}\right) \right] \mathbf{\hat{L}}_{\mathbf{N}}~,  \label{vtimesSkKA}
\end{eqnarray}%
and 
\begin{eqnarray}
\mathbf{\Omega }_{L}\times \mathbf{\hat{S}_{i}} &=&a_{3}\frac{\mu r}{L_{N}}%
\left\{ \left[ \frac{\sin \kappa _{i}}{\tan \alpha }\sin \left( \psi
_{p}-\psi _{i}\right) \sin \left( \psi _{p}+\chi _{p}\right) +\cos \kappa
_{i}\sin \chi _{p}\right] \mathbf{\hat{A}}_{\mathbf{N}}\right.  \notag \\
&&+\left[ \frac{\sin \kappa _{i}}{\tan \alpha }\cos \left( \psi _{p}-\psi
_{i}\right) \sin \left( \psi _{p}+\chi _{p}\right) -\cos \kappa _{i}\cos
\chi _{p}\right] \mathbf{\hat{Q}}_{\mathbf{N}}  \notag \\
&&\left. -\sin \kappa _{i}\sin \left( \psi _{p}+\chi _{p}-\psi _{i}\right) 
\mathbf{\hat{L}}_{\mathbf{N}}\right\} ~.  \label{OmLtimesSi}
\end{eqnarray}%
Eqs. (3)-(5) of \cite{IndepVar} give%
\begin{eqnarray}
S_{i} &=&\frac{G}{c}m^{2}\eta \nu ^{2i-3}\chi _{i}~,  \label{Si} \\
Q_{i} &=&-\frac{G^{2}}{c^{4}}w_{i}m^{2}\eta \nu ^{2i-3}\chi _{i}^{2}m_{i}~.
\label{Qi}
\end{eqnarray}

\subsection{Acceleration}

The general relativistic, SO, SS and QM contributions to the acceleration,
with the SO part given in the Newton-Wigner-Pryce spin supplementary
condition (NWP SSC) \cite{Kidder},\cite{quadrup}, by employing Eqs. (\ref{Si}%
)-(\ref{Qi}) are:%
\begin{equation}
\Delta \mathbf{a}=\mathbf{a}_{PN}+\mathbf{a}_{2PN}+\mathbf{a}_{SO}^{NWP}+%
\mathbf{a}_{SS}+\mathbf{a}_{QM}\ ,  \label{Deltaa}
\end{equation}%
with 
\begin{eqnarray}
\mathbf{a}_{PN} &=&{\frac{Gm}{c^{2}r^{2}}}\left\{ \left[ 2(2+\eta )\frac{Gm}{%
r}-(1+3\eta )v^{2}+{\frac{3}{2}}\eta \dot{r}^{2}\right] \mathbf{\hat{r}}%
+2(2-\eta )\dot{r}\mathbf{v}\right\} ~, \\
\mathbf{a}_{2PN} &=&-\frac{Gm}{c^{4}r^{2}}\Biggl\{\Bigl[\frac{3}{4}\left(
12+29\eta \right) \left( \frac{Gm}{r}\right) ^{2}+\eta \left( 3-4\eta
\right) v^{4}+\frac{15}{8}\eta \left( 1-3\eta \right) \dot{r}^{4}  \notag \\
&&-\frac{3}{2}\eta \left( 3-4\eta \right) \dot{r}^{2}v^{2}-\frac{\eta }{2}%
\left( 13-4\eta \right) \frac{Gm}{r}v^{2}-\left( 2+25\eta +2\eta ^{2}\right) 
\frac{Gm}{r}\dot{r}^{2}\Bigr]\mathbf{\hat{r}}  \notag \\
&&-\frac{1}{2}\left[ \eta \left( 15+4\eta \right) v^{2}-\left( 4+41\eta
+8\eta ^{2}\right) \frac{Gm}{r}-3\eta \left( 3+2\eta \right) \dot{r}^{2}%
\right] \dot{r}\mathbf{v}\Biggr\}~, \\
\mathbf{a}_{SO}^{NWP} &=&{\frac{G^{2}m^{2}\eta }{c^{3}r^{3}}}%
\sum_{k=1}^{2}\left( 4\nu ^{2k-3}+3\right) \chi _{k}\left\{ {\frac{3L_{N}}{%
2\mu r}}\left( \mathbf{\hat{L}}_{\mathbf{N}}\cdot \mathbf{\hat{S}}_{\mathbf{k%
}}\right) \mathbf{\hat{r}}-\left( \mathbf{v\times \hat{S}}_{\mathbf{k}%
}\right) +{\frac{3\dot{r}}{2}}\left( \mathbf{\hat{r}}\times \mathbf{\hat{S}}%
_{\mathbf{k}}\right) \right\} \\
\mathbf{a}_{SS} &=&-\frac{3G^{3}m^{3}\eta }{c^{4}r^{4}}\chi _{1}\chi
_{2}\left\{ \left[ \left( \mathbf{\hat{S}}_{\mathbf{1}}\cdot \mathbf{\hat{S}}%
_{\mathbf{2}}\right) -5\left( \mathbf{\hat{r}}\cdot \mathbf{\hat{S}}_{%
\mathbf{1}}\right) \left( \mathbf{\hat{r}}\cdot \mathbf{\hat{S}}_{\mathbf{2}%
}\right) \right] \mathbf{\hat{r}}+\left( \mathbf{\hat{r}}\cdot \mathbf{\hat{S%
}}_{\mathbf{2}}\right) \mathbf{\hat{S}}_{\mathbf{1}}+\left( \mathbf{\hat{r}}%
\cdot \mathbf{\hat{S}}_{\mathbf{1}}\right) \mathbf{\hat{S}}_{\mathbf{2}%
}\right\} \text{ }, \\
\mathbf{a}_{QM} &=&-\frac{3G^{3}m^{3}\eta }{2c^{4}r^{4}}\sum_{k=1}^{2}w_{k}%
\nu ^{2k-3}\chi _{k}^{2}\left\{ \left[ 1-5\left( \mathbf{\hat{r}\cdot \hat{S}%
}_{\mathbf{k}}\right) ^{2}\right] \mathbf{\hat{r}+}2\left( \mathbf{\hat{r}%
\cdot \hat{S}}_{\mathbf{k}}\right) \mathbf{\hat{S}}_{\mathbf{k}}\right\} 
\text{ }.
\end{eqnarray}%
After inserting Eqs. (\ref{rKA})-(\ref{vtimesSkKA}), the projections $a_{i}=$
$\Delta \mathbf{a\cdot f}_{(\mathbf{i})}$ with $\mathbf{f}_{(\mathbf{i})}=(%
\mathbf{\hat{A}}_{\mathbf{N}},\ \mathbf{\hat{Q}}_{\mathbf{N}},~\mathbf{\hat{L%
}}_{\mathbf{N}})$, they can be readily found. For explicit expressions we
also need 
\begin{eqnarray}
\mathbf{\hat{r}}\cdot \mathbf{\hat{S}_{k}} &=&\sin \kappa _{k}\cos \left(
\psi _{p}+\chi _{p}-\psi _{k}\right) ~,  \label{rdotSi} \\
\mathbf{\hat{S}}_{\mathbf{1}}\cdot \mathbf{\hat{S}}_{\mathbf{2}} &=&\cos
\kappa _{1}\cos \kappa _{2}+\sin \kappa _{1}\sin \kappa _{2}\cos \left( \psi
_{2}-\psi _{1}\right) ~.
\end{eqnarray}%
The acceleration components are

\begin{eqnarray}
a_{1} &=&a_{1}^{PN}+a_{1}^{2PN}+a_{1}^{SO}+a_{1}^{SS}+a_{1}^{QM}~,
\label{a1} \\
a_{1}^{PN} &=&{\frac{Gm}{c^{2}r^{2}}}\left\{ \left[ 2(2+\eta )\frac{Gm}{r}%
-(1+3\eta )v^{2}+{\frac{3}{2}}\eta \dot{r}^{2}\right] \cos \chi
_{p}-2(2-\eta )\dot{r}\frac{Gm\mu }{L_{N}}\sin \chi _{p}\right\} ~,  \notag
\\
a_{1}^{2PN} &=&-\frac{Gm}{c^{4}r^{2}}\Biggl\{\Bigl[\frac{3}{4}\left(
12+29\eta \right) \left( \frac{Gm}{r}\right) ^{2}+\eta \left( 3-4\eta
\right) v^{4}+\frac{15}{8}\eta \left( 1-3\eta \right) \dot{r}^{4}  \notag \\
&&-\frac{3}{2}\eta \left( 3-4\eta \right) \dot{r}^{2}v^{2}-\frac{\eta }{2}%
\left( 13-4\eta \right) \frac{Gm}{r}v^{2}-\left( 2+25\eta +2\eta ^{2}\right) 
\frac{Gm}{r}\dot{r}^{2}\Bigr]\cos \chi _{p}  \notag \\
&&+\left[ \eta \left( 15+4\eta \right) v^{2}-\left( 4+41\eta +8\eta
^{2}\right) \frac{Gm}{r}-3\eta \left( 3+2\eta \right) \dot{r}^{2}\right] 
\frac{Gm\mu \dot{r}}{2L_{N}}\sin \chi _{p}\Biggr\}~,  \notag \\
a_{1}^{SO} &=&{\frac{G^{2}m^{2}\eta }{c^{3}r^{3}}}\left[ \left( {\frac{3L_{N}%
}{2\mu r}}-\frac{Gm\mu }{L_{N}}\right) \cos \chi _{p}+{\frac{3\dot{r}}{2}}%
\sin \chi _{p}-\frac{A_{N}}{L_{N}}\right] \sum_{k=1}^{2}\left( 4\nu
^{2k-3}+3\right) \chi _{k}\cos \kappa _{k}~,  \notag \\
a_{1}^{SS} &=&-\frac{3G^{3}m^{3}\eta }{c^{4}r^{4}}\chi _{1}\chi _{2}\Bigl\{%
\cos \kappa _{1}\cos \kappa _{2}\cos \chi _{p}+\sin \kappa _{1}\sin \kappa
_{2}  \notag \\
&&\times \bigl[\left[ \cos \left( \psi _{2}-\psi _{1}\right) -5\cos \left(
\psi _{p}+\chi _{p}-\psi _{1}\right) \cos \left( \psi _{p}+\chi _{p}-\psi
_{2}\right) \right] \cos \chi _{p}  \notag \\
&&+\cos \left( \psi _{p}+\chi _{p}-\psi _{2}\right) \cos \left( \psi
_{p}-\psi _{1}\right) +\cos \left( \psi _{p}+\chi _{p}-\psi _{1}\right) \cos
\left( \psi _{p}-\psi _{2}\right) \bigr]\Bigr\}~,  \notag \\
a_{1}^{QM} &=&-\frac{3G^{3}m^{3}\eta }{2c^{4}r^{4}}\sum_{k=1}^{2}w_{k}\nu
^{2k-3}\chi _{k}^{2}\Bigl\{\cos \chi _{p}-\sin ^{2}\kappa _{k}\cos \left(
\psi _{p}+\chi _{p}-\psi _{k}\right)   \notag \\
&&\times \left[ 5\cos \chi _{p}\cos \left( \psi _{p}+\chi _{p}-\psi
_{k}\right) \mathbf{-}2\cos \left( \psi _{p}-\psi _{k}\right) \right] \Bigr\}%
~,  \notag
\end{eqnarray}%
$\allowbreak $%
\begin{eqnarray}
a_{2} &=&a_{2}^{PN}+a_{2}^{2PN}+a_{2}^{SO}+a_{2}^{SS}+a_{2}^{QM}~,
\label{a2} \\
a_{2}^{PN} &=&{\frac{Gm}{c^{2}r^{2}}}\left\{ \left[ 2(2+\eta )\frac{Gm}{r}%
-(1+3\eta )v^{2}+{\frac{3}{2}}\eta \dot{r}^{2}\right] \sin \chi
_{p}+2(2-\eta )\dot{r}\frac{Gm\mu }{L_{N}}\left( \cos \chi _{p}+\frac{A_{N}}{%
Gm\mu }\right) \right\}   \notag \\
a_{2}^{2PN} &=&-\frac{Gm}{c^{4}r^{2}}\Biggl\{\Bigl[\frac{3}{4}\left(
12+29\eta \right) \left( \frac{Gm}{r}\right) ^{2}+\eta \left( 3-4\eta
\right) v^{4}+\frac{15}{8}\eta \left( 1-3\eta \right) \dot{r}^{4}  \notag \\
&&-\frac{3}{2}\eta \left( 3-4\eta \right) \dot{r}^{2}v^{2}-\frac{\eta }{2}%
\left( 13-4\eta \right) \frac{Gm}{r}v^{2}-\left( 2+25\eta +2\eta ^{2}\right) 
\frac{Gm}{r}\dot{r}^{2}\Bigr]\sin \chi _{p}  \notag \\
&&-\left[ \eta \left( 15+4\eta \right) v^{2}-\left( 4+41\eta +8\eta
^{2}\right) \frac{Gm}{r}-3\eta \left( 3+2\eta \right) \dot{r}^{2}\right] 
\frac{Gm\mu \dot{r}}{2L_{N}}\left( \cos \chi _{p}+\frac{A_{N}}{Gm\mu }%
\right) \Biggr\}~,  \notag \\
a_{2}^{SO} &=&{\frac{G^{2}m^{2}\eta }{c^{3}r^{3}}}\left[ \left( {\frac{3L_{N}%
}{2\mu r}}-\frac{Gm\mu }{L_{N}}\right) \sin \chi _{p}-{\frac{3\dot{r}}{2}}%
\cos \chi _{p}\right] \sum_{k=1}^{2}\left( 4\nu ^{2k-3}+3\right) \chi
_{k}\cos \kappa _{k}~,  \notag \\
a_{2}^{SS} &=&-\frac{3G^{3}m^{3}\eta }{c^{4}r^{4}}\chi _{1}\chi _{2}\Bigl\{%
\cos \kappa _{1}\cos \kappa _{2}\sin \chi _{p}+\sin \kappa _{1}\sin \kappa
_{2}  \notag \\
&&\times \bigl[\left[ \cos \left( \psi _{2}-\psi _{1}\right) -5\cos \left(
\psi _{p}+\chi _{p}-\psi _{1}\right) \cos \left( \psi _{p}+\chi _{p}-\psi
_{2}\right) \right] \sin \chi _{p}  \notag \\
&&-\cos \left( \psi _{p}+\chi _{p}-\psi _{2}\right) \sin \left( \psi
_{p}-\psi _{1}\right) -\cos \left( \psi _{p}+\chi _{p}-\psi _{1}\right) \sin
\left( \psi _{p}-\psi _{2}\right) \bigr]\Bigr\}~,  \notag \\
a_{2}^{QM} &=&-\frac{3G^{3}m^{3}\eta }{2c^{4}r^{4}}\sum_{k=1}^{2}w_{k}\nu
^{2k-3}\chi _{k}^{2}\Bigl\{\sin \chi _{p}-\sin ^{2}\kappa _{k}\cos \left(
\psi _{p}+\chi _{p}-\psi _{k}\right)   \notag \\
&&\times \left[ 5\sin \chi _{p}\cos \left( \psi _{p}+\chi _{p}-\psi
_{k}\right) \mathbf{+}2\sin \left( \psi _{p}-\psi _{k}\right) \right] \Bigr\}%
~,  \notag
\end{eqnarray}%
and%
\begin{eqnarray}
a_{3} &=&a_{3}^{SO}+a_{3}^{SS}+a_{3}^{QM}~,  \label{a3} \\
a_{3}^{SO} &=&{\frac{G^{2}m^{2}\eta }{c^{3}r^{3}}}\sum_{k=1}^{2}\left( 4\nu
^{2k-3}+3\right) \chi _{k}\sin \kappa _{k}  \notag \\
&&\times \left\{ \frac{Gm\mu }{L_{N}}\left[ \cos \left( \psi _{p}+\chi
_{p}-\psi _{k}\right) +\frac{A_{N}}{Gm\mu }\cos \left( \psi _{p}-\psi
_{k}\right) \right] -{\frac{3\dot{r}}{2}}\sin \left( \psi _{p}+\chi
_{p}-\psi _{k}\right) \right\} ~,  \notag \\
a_{3}^{SS} &=&-\frac{3G^{3}m^{3}\eta }{c^{4}r^{4}}\chi _{1}\chi _{2}\left[
\cos \kappa _{1}\sin \kappa _{2}\cos \left( \psi _{p}+\chi _{p}-\psi
_{2}\right) +\cos \kappa _{2}\sin \kappa _{1}\cos \left( \psi _{p}+\chi
_{p}-\psi _{1}\right) \right] ~,  \notag \\
a_{3}^{QM} &=&-\frac{3G^{3}m^{3}\eta }{2c^{4}r^{4}}\sum_{k=1}^{2}w_{k}\nu
^{2k-3}\chi _{k}^{2}\sin 2\kappa _{k}\cos \left( \psi _{p}+\chi _{p}-\psi
_{k}\right) ~.  \notag
\end{eqnarray}

In the above expressions we still need to employ Eqs. (20)-(21) and (23) of 
\cite{IndepVar} 
\begin{eqnarray}
r &=&\frac{L_{N}^{2}}{\mu \left( Gm\mu +A_{N}\cos \chi _{p}\right) }~,
\label{truer} \\
\dot{r} &=&\frac{A_{N}}{L_{N}}\sin \chi _{p}~,  \label{truerdot} \\
v^{2} &=&\frac{\left( Gm\mu \right) ^{2}+A_{N}^{2}+2Gm\mu A_{N}\cos \chi _{p}%
}{L_{N}^{2}}~,  \label{truev2}
\end{eqnarray}%
in order to rewrite $r$, $\dot{r}$ and $v^{2}$ in terms of the chosen
dynamical variables.

Also, as $L_{N}$ is not among the chosen independent variables, we need to
express it in terms of them. For this, first we give the SO part of the
orbital angular momentum in the NWP SSC: 
\begin{eqnarray}
\mathbf{L}_{\mathbf{SO}}^{NWP} &=&\frac{G\mu }{2c^{2}r}\sum_{k=1}^{2}\left(
4+3\nu ^{3-2i}\right) S_{i}\left[ \mathbf{\hat{r}}\times \left( \mathbf{\hat{%
r}}\times \mathbf{\hat{S}_{i}}\right) \right]  \notag \\
&=&\frac{G^{2}m^{3}}{4c^{3}r}\eta ^{2}\sum_{i=1}^{2}\left( 4\nu
^{2i-3}+3\right) \chi _{i}\left\{ \sin \kappa _{i}\left[ \allowbreak \cos
\left( 2\chi _{p}+\psi _{p}-\psi _{i}\right) -\cos \left( \psi _{p}-\psi
_{i}\right) \right] \mathbf{\hat{A}}_{\mathbf{N}}\right.  \notag \\
&&\left. +\sin \kappa _{i}\left[ \sin \left( 2\chi _{p}+\psi _{p}-\psi
_{i}\right) +\sin \left( \psi _{p}-\psi _{i}\right) \right] \mathbf{\hat{Q}}%
_{\mathbf{N}}-2\cos \kappa _{i}\mathbf{\hat{L}}_{\mathbf{N}}\right\} ~.
\label{LSONWP}
\end{eqnarray}%
As expected its order is 
\begin{equation}
\mathcal{O}\left( \frac{\mathbf{L}_{\mathbf{SO}}^{NWP}}{L_{N}}\right) =%
\mathcal{O}\left( \varepsilon ^{3/2}\right) \mathcal{O}\left( \eta \right) 
\mathcal{O}\left( 1,\nu ^{2i-3}\right) \chi _{i}~.  \label{epordo2}
\end{equation}%
The total angular momentum $\mathbf{J=L+S}_{\mathbf{1}}\mathbf{+S}_{\mathbf{2%
}}$ in the system $\mathcal{K}_{A}$ then becomes, using Eqs. (\ref{SiKA})
and (\ref{LSONWP}):%
\begin{gather}
J\mathbf{\hat{J}}=\frac{Gm^{2}}{c}\eta \sum_{i=1}^{2}\chi _{i}\sin \kappa
_{i}  \notag \\
\times \left\{ \left[ \nu ^{2i-3}\cos \left( \psi _{p}-\psi _{i}\right) -%
\frac{Gm}{4c^{2}r}\eta \left( 4\nu ^{2i-3}+3\right) \left[ \cos \left( \psi
_{p}-\psi _{i}\right) -\cos \left( 2\chi _{p}+\psi _{p}-\psi _{i}\right) %
\right] \right] \mathbf{\hat{A}}_{\mathbf{N}}\right.  \notag \\
\left. -\left[ \nu ^{2i-3}\sin \left( \psi _{p}-\psi _{i}\right) -\frac{Gm}{%
4c^{2}r}\eta \left( 4\nu ^{2i-3}+3\right) \left[ \sin \left( \psi _{p}-\psi
_{i}\right) +\sin \left( 2\chi _{p}+\psi _{p}-\psi _{i}\right) \right] %
\right] \mathbf{\hat{Q}}_{\mathbf{N}}\right\}  \notag \\
+\left\{ L_{N}\left( 1+\epsilon _{PN}+\epsilon _{2PN}\right) +\frac{Gm^{2}}{c%
}\eta \sum_{i=1}^{2}\left[ \nu ^{2i-3}-\frac{Gm}{2c^{2}r}\eta \left( 4\nu
^{2i-3}+3\right) \right] \chi _{i}\cos \kappa _{i}\right\} \mathbf{\hat{L}}_{%
\mathbf{N}}~,  \label{JdirKA}
\end{gather}%
Here $\epsilon _{PN}$ and $\epsilon _{2PN}$ are given by Eqs. (39)-(40) of 
\cite{IndepVar}. The projections along the basis vectors $\mathbf{\hat{l}}$, 
$\mathbf{\hat{m}}$, $\mathbf{\hat{L}}_{\mathbf{N}}$ of the $\mathcal{K}_{L}$
system are\footnote{%
These are the equations in the NWP SSC corresponding to Eqs. (46)-(48) of 
\cite{IndepVar}, which were written in the covariant SSC.}%
\begin{gather}
0=\sum_{i=1}^{2}\chi _{i}\sin \kappa _{i}\left[ \nu ^{2i-3}\cos \psi _{i}+%
\frac{Gm}{4c^{2}r}\eta \left( 4\nu ^{2i-3}+3\right) \left[ \cos \left( 2\chi
_{p}+2\psi _{p}-\psi _{i}\right) -\cos \psi _{i}\right] \right] \ ,
\label{proj1} \\
\frac{cJ\sin \alpha }{Gm^{2}\eta }=\sum_{i=1}^{2}\chi _{i}\sin \kappa _{i}%
\left[ \nu ^{2i-3}\sin \psi _{i}+\frac{Gm}{4c^{2}r}\eta \left( 4\nu
^{2i-3}+3\right) \left[ \allowbreak \sin \left( 2\chi _{p}+2\psi _{p}-\psi
_{i}\right) -\sin \psi _{i}\right] \right] \ ,  \label{proj2} \\
J\cos \alpha =L_{N}\left( 1+\epsilon _{PN}+\epsilon _{2PN}\right) +\frac{%
Gm^{2}}{c}\eta \sum_{i=1}^{2}\left[ \nu ^{2i-3}-\frac{Gm}{2c^{2}r}\eta
\left( 4\nu ^{2i-3}+3\right) \right] \chi _{i}\cos \kappa _{i}\ .
\label{proj3}
\end{gather}%
The last equation enables us to express $L_{N}$ to 2PN accuracy in terms of
the chosen independent variables:%
\begin{eqnarray}
L_{N} &=&J\left( 1-\epsilon _{PN}-\epsilon _{2PN}+\epsilon _{PN}^{2}\right)
\cos \alpha  \notag \\
&&-\frac{Gm^{2}}{c}\eta \sum_{i=1}^{2}\left[ \left( 1-\epsilon _{PN}\right)
\nu ^{2i-3}-\frac{Gm}{2c^{2}r}\eta \left( 4\nu ^{2i-3}+3\right) \right] \chi
_{i}\cos \kappa _{i}\ .  \label{LNtaylor}
\end{eqnarray}%
We also give the series expansion of its reciprocal:

\begin{eqnarray}
\frac{1}{L_{N}} &=&\frac{1+\epsilon _{PN}+\epsilon _{2PN}}{J\cos \alpha }%
+\left( \frac{Gm^{2}}{cJ\cos \alpha }\right) \frac{\eta }{J\cos \alpha }%
\left[ \left( 1+\epsilon _{PN}\right) \chi _{\nu }-\frac{Gm}{2c^{2}r}\eta
\left( 4\chi _{\nu }+3\chi _{+}\right) \right]  \notag \\
&&+\left( \frac{Gm^{2}}{cJ\cos \alpha }\right) ^{2}\frac{\eta ^{2}}{J\cos
\alpha }\left[ \left( 1+\epsilon _{PN}\right) \chi _{\nu }-\frac{Gm}{c^{2}r}%
\eta \left( 4\chi _{\nu }+3\chi _{+}\right) \right] \chi _{\nu }  \notag \\
&&+\left( \frac{Gm^{2}}{cJ\cos \alpha }\right) ^{3}\frac{\eta ^{3}}{J\cos
\alpha }\chi _{\nu }^{3}+\left( \frac{Gm^{2}}{cJ\cos \alpha }\right) ^{4}%
\frac{\eta ^{4}}{J\cos \alpha }\chi _{\nu }^{4}\ ,
\end{eqnarray}%
where we employed the notations 
\begin{eqnarray}
\chi _{+} &=&\sum_{i=1}^{2}\chi _{i}\cos \kappa _{i}=\chi _{1}\cos \kappa
_{1}+\chi _{2}\cos \kappa _{2}~,  \notag \\
\chi _{\nu } &=&\sum_{i=1}^{2}\nu ^{2i-3}\chi _{i}\cos \kappa _{i}=\nu
^{-1}\chi _{1}\cos \kappa _{1}+\nu \chi _{2}\cos \kappa _{2}~.
\label{chipchinu}
\end{eqnarray}%
Note that the 2PN contribution of $1/L_{N}$ is rather messy (fourth rank in
the spins), nevertheless for our purposes we need it only to 1PN accuracy
(it enters only in PN terms or higher, and the desired accuracy is 2PN).

We also give here the detailed expression in terms of orbital elements of $%
\epsilon _{PN}$, which is necessary at this accuracy: 
\begin{eqnarray}
\epsilon _{PN} &=&\frac{1-3\eta }{2}\left( \frac{v}{c}\right) ^{2}+\left(
3+\eta \right) \frac{Gm}{c^{2}r}  \notag \\
&=&\frac{\left( 7-\eta \right) \left( Gm\mu \right) ^{2}+\left( 1-3\eta
\right) A_{N}^{2}+4\left( 2-\eta \right) Gm\mu A_{N}\cos \chi _{p}}{%
2c^{2}L_{N}^{2}}  \notag \\
&=&\frac{\left( Gm\mu \right) ^{2}}{2c^{2}J^{2}\cos ^{2}\alpha }\left[
\left( 1-3\eta \right) e_{r}^{2}+4\left( 2-\eta \right) e_{r}\cos \chi
_{p}+\left( 7-\eta \right) \right] ~.
\end{eqnarray}

\subsection{Spin angular velocity}

The spin undergoes a pure precession, therefore its magnitude is unchanged,
while its direction changes as 
\begin{equation}
\frac{d}{dt}\mathbf{\hat{S}_{i}=\Omega }_{\mathbf{i}}\times \mathbf{\hat{S}%
_{i}~,}
\end{equation}%
where, after employing Eqs. (\ref{Si})-(\ref{Qi}), (\ref{rKA}), (\ref{SiKA}%
), and (\ref{rdotSi}) in Eqs. (56) of \cite{IndepVar} the angular velocity
vector is found as 
\begin{eqnarray}
\mathbf{\Omega }_{\mathbf{i}} &=&\mathbf{\Omega }_{\mathbf{i}}^{SO}+\mathbf{%
\Omega }_{\mathbf{i}}^{SS}+\mathbf{\Omega }_{\mathbf{i}}^{QM}~,
\label{Omivec} \\
\mathbf{\Omega }_{\mathbf{i}}^{SO} &=&{\frac{G\left( 4+3\nu ^{3-2i}\right) }{%
2c^{2}r^{3}}L}_{N}\mathbf{\hat{L}_{N}}~,  \notag \\
\mathbf{\Omega }_{\mathbf{i}}^{SS} &=&\frac{G^{2}m^{2}\eta }{2c^{3}r^{3}}\nu
^{2j-3}\chi _{j}\left[ \sin \kappa _{j}\left\{ \left[ 3\cos \left( \psi
_{p}+\chi _{p}-\psi _{j}\right) \cos \chi _{p}\mathbf{-}\cos \left( \psi
_{p}-\psi _{j}\right) \right] \mathbf{\hat{A}}_{\mathbf{N}}\right. \right. 
\notag \\
&&\left. \left. +\left[ 3\cos \left( \psi _{p}+\chi _{p}-\psi _{j}\right)
\sin \chi _{p}\mathbf{+}\sin \left( \psi _{p}-\psi _{j}\right) \right] 
\mathbf{\hat{Q}}_{\mathbf{N}}\right\} -\cos \kappa _{j}\mathbf{\hat{L}}_{%
\mathbf{N}}\right] ~,  \notag \\
\mathbf{\Omega }_{\mathbf{i}}^{QM} &=&\frac{G^{2}m^{2}\eta }{2c^{3}r^{3}}%
3w_{i}\chi _{i}\sin \kappa _{i}\cos \left( \psi _{p}+\chi _{p}-\psi
_{i}\right) \left( \cos \chi _{p}\mathbf{\hat{A}}_{\mathbf{N}}+\sin \chi _{p}%
\mathbf{\hat{Q}}_{\mathbf{N}}\right) \mathbf{~,}  \notag
\end{eqnarray}%
with $j\neq i$\thinspace . Their PN\ order is%
\begin{eqnarray}
\mathcal{O}\left( \mathbf{\Omega }_{\mathbf{i}}^{SO}\right) &=&\mathcal{O}%
\left( \varepsilon \right) \mathcal{O}\left( 1,\nu ^{3-2i}\right) \mathcal{O}%
(T^{-1})~,  \notag \\
\mathcal{O}\left( \mathbf{\Omega }_{\mathbf{i}}^{SS}\right) &=&\mathcal{O}%
\left( \varepsilon ^{3/2}\right) \mathcal{O}\left( \eta \right) \mathcal{O}%
\left( \nu ^{2i-3}\right) \chi _{j}\mathcal{O}(T^{-1})~,  \notag \\
\mathcal{O}\left( \mathbf{\Omega }_{\mathbf{i}}^{QM}\right) &=&\mathcal{O}%
\left( \varepsilon ^{3/2}\right) \mathcal{O}\left( \eta \right) w_{i}\chi
_{i}\mathcal{O}(T^{-1})~.  \label{LS1S2estimates}
\end{eqnarray}

The projections employed in the main text are%
\begin{eqnarray}
\mathbf{\Omega }_{\mathbf{i}}\cdot \mathbf{\hat{A}}_{\mathbf{N}} &=&\frac{%
G^{2}m^{2}\eta }{2c^{3}r^{3}}\left\{ \nu ^{2j-3}\chi _{j}\sin \kappa _{j}%
\left[ 3\cos \left( \psi _{p}+2\chi _{p}-\psi _{j}\right) +\cos \left( \psi
_{p}-\psi _{j}\right) \right] \right.  \notag \\
&&\left. +3w_{i}\chi _{i}\sin \kappa _{i}\left[ \cos \left( \psi _{p}+2\chi
_{p}-\psi _{i}\right) +\cos \left( \psi _{p}-\psi _{i}\right) \right]
\right\} ~,  \label{Omiproj1} \\
\mathbf{\Omega }_{\mathbf{i}}\cdot \mathbf{\hat{Q}}_{\mathbf{N}} &=&\frac{%
G^{2}m^{2}\eta }{2c^{3}r^{3}}\left\{ \nu ^{2j-3}\chi _{j}\sin \kappa _{j}%
\left[ 3\sin \left( \psi _{p}+2\chi _{p}-\psi _{j}\right) -\sin \left( \psi
_{p}-\psi _{j}\right) \right] \right.  \notag \\
&&\left. +3w_{i}\chi _{i}\sin \kappa _{i}\left[ \sin \left( \psi _{p}+2\chi
_{p}-\psi _{i}\right) -\sin \left( \psi _{p}-\psi _{i}\right) \right]
\right\} \mathbf{~,}  \label{Omiproj2} \\
\mathbf{\Omega }_{\mathbf{i}}\cdot \mathbf{\hat{L}}_{\mathbf{N}} &=&{\frac{%
G\left( 4+3\nu ^{3-2i}\right) }{2c^{2}r^{3}}}J\cos \alpha -\frac{%
G^{2}m^{2}\eta }{2c^{3}r^{3}}\left[ \left( 4\nu ^{2i-3}+3\right) \chi
_{i}\cos \kappa _{i}+\nu ^{2j-3}\left( 5+3\nu ^{3-2i}\right) \chi _{j}\cos
\kappa _{j}\right] ~.  \label{Omiproj3}
\end{eqnarray}%
We also need 
\begin{eqnarray}
\mathbf{\Omega }_{\mathbf{i}}\times \mathbf{\hat{S}_{i}} &=&\left[ \left( 
\mathbf{\Omega }_{\mathbf{i}}\cdot \mathbf{\hat{L}}_{\mathbf{N}}\right) \sin
\kappa _{i}\sin \left( \psi _{p}-\psi _{i}\right) +\left( \mathbf{\Omega }_{%
\mathbf{i}}\cdot \mathbf{\hat{Q}}_{\mathbf{N}}\right) \cos \kappa _{i}\right]
\mathbf{\hat{A}}_{\mathbf{N}}  \notag \\
&&+\left[ \left( \mathbf{\Omega }_{\mathbf{i}}\cdot \mathbf{\hat{L}}_{%
\mathbf{N}}\right) \sin \kappa _{i}\cos \left( \psi _{p}-\psi _{i}\right)
-\left( \mathbf{\Omega }_{\mathbf{i}}\cdot \mathbf{\hat{A}}_{\mathbf{N}%
}\right) \cos \kappa _{i}\right] \mathbf{\hat{Q}}_{\mathbf{N}}  \notag \\
&&-\sin \kappa _{i}\left[ \left( \mathbf{\Omega }_{\mathbf{i}}\cdot \mathbf{%
\hat{A}}_{\mathbf{N}}\right) \sin \left( \psi _{p}-\psi _{i}\right) +\left( 
\mathbf{\Omega }_{\mathbf{i}}\cdot \mathbf{\hat{Q}}_{\mathbf{N}}\right) \cos
\left( \psi _{p}-\psi _{i}\right) \right] \mathbf{\hat{L}}_{\mathbf{N}}~.
\label{OmitimesSi}
\end{eqnarray}

\end{document}